\documentstyle[12pt]{article}

\def\df #1. #2\par{\medbreak
  \noindent{{\tt {\bf Definition #1.}}\enspace}{\sl#2\par}%
  \ifdim\lastskip<\medskipamount \removelastskip\penalty55\medskip\fi}

\def\theorem #1. #2\par{\medbreak
  \noindent{\tt {\bf Theorem #1.}\enspace}{\sl#2\par}%
  \ifdim\lastskip<\medskipamount \removelastskip\penalty55\medskip\fi}

\def\lemma #1. #2\par{\medbreak
  \noindent{\tt {\bf Lemma #1.}\enspace}{\sl#2\par}%
  \ifdim\lastskip<\medskipamount \removelastskip\penalty55\medskip\fi}

\def\proof{\medbreak\noindent{\bf Proof}}

\newcommand{\s}[1]{\Lambda_{#1}}
\newcommand{\num}[1]{\par\medskip\noindent{\large\bf #1.\,\,}}

\def\la{\Lambda}
\def\lak{\Lambda^{\K}}

\def\vol{{\ell }}

\def\cl{{\cal C}\!\ell }
\def\fin{\hbox{$\bullet$}\medskip}
\def\D{{\cal D}}
\def\M{{\cal M}}
\def\L{{\cal L}}
\def\V{{\cal V}}
\def\K{{\cal K}}
\def\R{{\cal R}}
\def\C{{\cal C}}

\def\be{\begin{equation}}
\def\ee{\end{equation}}
\def\tr{{\rm Tr}}
\def\ww{\wedge\ldots\wedge}

\def\det{{\rm det}}
\def\even{{\rm ev}}
\def\odd{{\rm od}}

\def\kfac{\frac{1}{k!}}

\def\Spin{{\rm Spin}}
\def\com{{\rm Com}}
\def\exp{{\rm exp}}
\def\diag{{\rm diag}}

\begin{document}

\title{Dirac-type tensor equations with non-Abelian gauge symmetries on
pseudo-Riemannian space}

\author{N.G.Marchuk \thanks{Research supported by the Russian Foundation
for Basic Research grants 00-01-00224,  00-15-96073.}}

\begin{abstract}
We suggest a so-called Dirac type tensor equation with nonabelian gauge
symmetry on pseudo-Riemannian space. This equation reproduce some
of the properties of spinor Dirac equation. A geometrical interpretation
of results in terms of Riemannian geometry is given.
\end{abstract}

\maketitle


\vskip 1cm

Steklov Mathematical Institute, Gubkina st.8, Moscow 119991, Russia

nmarchuk@mi.ras.ru,   www.orc.ru/~nmarchuk


\tableofcontents

\bigskip

In the previous paper \cite{Marchuk:Cimento1}, developing results of P.~Dirac \cite{Dirac},
D.~D.~Ivanenko, L.~D.~Landau \cite{Ivanenko}, E.~K\"ahler \cite{Kahler},
F.~G\"ursey \cite{Gursey}, D.~Hestenes \cite{Hestenes}, \cite{Hestenes1},
we consider  the
so-called Dirac type tensor equation which reproduce some of the properties of the
spinor Dirac equation for an electron. Now we generalize the Dirac type tensor equation
in two ways.

Firstly we present the Dirac type tensor equation with nonabelian gauge symmetry. We
use unitary gauge Lie groups that are subgroups of the group $U(4)$, in particular
$U(1)$, $U(1)\times SU(2)$, $SU(3)$ -- main groups of the Standard Model. Note
that in case of the gauge group $SU(3)$ the Dirac type tensor equation has 16 complex
valued components of wave function. But the Dirac equation for chromospinor has only
12 complex valued components. Evidently these two systems of equations are not coincide.

Secondly we generalize all results on pseudo-Riemannian space (in \cite{Marchuk:Cimento1}
equations
are considered in Minkowski space). A method of generalization was suggested
in \cite{Marchuk:Cimento}. A key role in this method play the tensor $B_\mu$.
Now we have found explicit formulas for $B_\mu$ via the components of metric tensor
$g_{\mu\nu}$ and its first derivatives. In the section 7 we give an interpretation
of our results in terms of Riemannian geometry.


\section{Basic definitions.}
More details about the following definitions can be found in \cite{Marchuk:Cimento}.

\num{1}
Let $\M$ be a four dimensional differentiable manifolds with a local
system of coordinates $x^\mu$. Greek indices run over (1,2,3,4).
Summation convention over repeating indices is assumed.
Suppose that there is a smooth
twice covariant tensor field (metric tensor) with components
$g_{\mu\nu}=g_{\mu\nu}(x)$, $x\in\M$ such that
\begin{itemize}
\item $g_{\mu\nu}=g_{\nu\mu}$;
\item $g = \det\|g_{\mu\nu}\|<0$;
\item The signature of the matrix $\|g_{\mu\nu}\|$ is equal to $-2$.
\end{itemize}

The matrix $\|g^{\mu\nu}\|$ composed from contravariant components of
the metric tensor is the inverse matrix to $\|g_{\mu\nu}\|$.
The full set of $\{\M,g_{\mu\nu}\}$ is called {\sl a
pseudo-Riemannian space}  and  denoted by $\V$.

\num{2}Let $\s k$ be the sets of exterior differential
forms of rank $k=0,1,2,3,4$ on $\V$ (covariant antisymmetric tensor fields) and
$$
\la=\s 0\oplus\ldots\oplus\s 4=\s \even\oplus\s \odd,
\quad\s \even=\s 0\oplus\s 2\oplus\s 4,\quad
\s \odd=\s 1\oplus\s 3.
$$
Elements of $\la$ are called (nonhomogeneous) {\sl differential forms}
and elements of $\s k$ are called {\sl $k$-forms} or differential forms
of rank $k$. Elements of $\s{\even}$ and $\s\odd$ are called {\sl even}
and {\sl odd} forms respectively.
 The set of smooth scalar functions on $\V$ (invariants) is
identified with the set of $0$-forms $\s 0$. A $k$-form $U\in\s k$ can
be written as
\begin{equation}
U=\kfac u_{\nu_1\ldots \nu_k}dx^{\nu_1}\ww dx^{\nu_k}=
\sum_{\mu_1<\cdots<\mu_k} u_{\mu_1\ldots \mu_k}dx^{\mu_1}\ww dx^{\mu_k},
\label{k-form}
\end{equation}
where  the smooth functions $u_{\nu_1\ldots \nu_k}=u_{\nu_1\ldots \nu_k}(x)$
are real valued
components of a covariant antisymmetric
($u_{\nu_1\ldots \nu_k}=u_{[\nu_1\ldots \nu_k]}$) tensor field.

Let $\s k^\C$, $\la^\C$, $\s \even^\C$, $\s \odd^\C$ be corresponding
sets of complex valued differential forms and $\K$ be a field
of complex numbers $\C$ or real numbers $\R$ ($\la^\R=\la$).
Differential forms from $\la^\K$ can be written as linear combinations of
the 16 basis differential forms
\begin{equation}
1,dx^\mu,dx^{\mu_1}\wedge dx^{\mu_2},\ldots,dx^{0}\wedge\ldots\wedge
dx^3,
\quad
\mu_1<\mu_2<\ldots
\label{basis}
\end{equation}
with coefficients from $C^\infty(\V,\K)$ (smooth functions that map
$\V$ into $\K$).
The exterior product of differential forms is defined in the
usual way.
If $U\in\s{r}^\K,V\in\s{s}^\K$, then
$$
U\wedge V=(-1)^{rs}V\wedge U\in\s{r+s}^\K.
$$

\num{3}Consider the Hodge star operator
$\star\,:\,\s{k}^\K\to\s{4-k}^\K$. If $U\in\s k$
has the form (\ref{k-form}), then
$$
\star U=
\frac{1}{k!(4-k)!}\sqrt{-g}\,\varepsilon_{\mu_1\ldots\mu_4}u^{\mu_1\ldots\mu_k}
dx^{\mu_{k+1}}\ww dx^{\mu_4},
$$
where
$u^{\mu_1\ldots\mu_k}=g^{\mu_1\nu_1}\ldots g^{\mu_k\nu_k}u_{\nu_1\ldots\nu_k}
$,
$\varepsilon_{\mu_1\ldots\mu_4}$ is the sign of the permutation
$(\mu_1\ldots\mu_4)$, and $\varepsilon_{0123}=1$. It is easy to prove
that for $U\in\s{k}^\K$
$$
\star(\star U)=(-1)^{k+1}U.
$$

The form $\star U$ is a covariant antisymmetric tensor with respect to
changes of coordinates with positive Jacobian.

\num{4} Let us define the {\sl volume form}
$$
\vol  = \sqrt{-g}\, dx^0\wedge dx^1\wedge dx^2\wedge dx^3=\frac{\sqrt{-g}}{4!}
\epsilon_{\mu_1\ldots\mu_4}dx^{\mu_1}\ww dx^{\mu_4}.
$$
We have
$$
(\vol )^2=-1,\quad \stackrel{k}{U}\vol =(-1)^k \vol  \stackrel{k}{U},\quad\hbox{for}\quad
\stackrel{k}{U}\in\Lambda_k
$$
that means $\vol $ commutes with all even forms and anticommutes with all odd forms
with respect to Clifford product (see below).

\num{5}Further on we consider the bilinear operator
$\com\,:\,\s 2\times\s 2\to\s 2$ such that
\begin{eqnarray*}
&&\com(\frac{1}{2}a_{\mu_1\mu_2}dx^{\mu_1}\wedge dx^{\mu_2},
\frac{1}{2}b_{\nu_1\nu_2}dx^{\nu_1}\wedge dx^{\nu_2})=
\frac{1}{2}a_{\mu_1\mu_2}b_{\nu_1\nu_2}(-g^{\mu_1 \nu_1}dx^{\mu_2}\wedge dx^{\nu_2}\\
&&-g^{\mu_2 \nu_2}dx^{\mu_1}\wedge dx^{\nu_1}
+g^{\mu_1 \nu_2}dx^{\mu_2}\wedge dx^{\nu_1}+g^{\mu_2
\nu_1}dx^{\mu_1}\wedge dx^{\nu_2}),
\end{eqnarray*}
where $a_{\mu_1\mu_2}=a_{[\mu_1\mu_2]}$,
$b_{\nu_1\nu_2}=b_{[\nu_1\nu_2]}$.
Evidently, $\com(U,V)=-\com(V,U)$.

\num{6}Now we define the Clifford product of differential forms with the aid of
the following formulas:
\begin{eqnarray*}
\stackrel{0}{U}\stackrel{k}{V}&=&\stackrel{k}{V}\stackrel{0}{U}=\stackrel{0}{U}\wedge\stackrel{k}{V}=\stackrel{k}{V}\wedge\stackrel{0}{U},\\
\stackrel{1}{U}\stackrel{k}{V}&=&\stackrel{1}{U} \wedge  \stackrel{k}{V}-\star (\stackrel{1}{U} \wedge  \star \stackrel{k}{V}),\\
\stackrel{k}{U}\stackrel{1}{V}&=&\stackrel{k}{U} \wedge  \stackrel{1}{V}+\star ( \star \stackrel{k}{U} \wedge \stackrel{1}{V}),\\
\stackrel{2}{U}\stackrel{2}{V}&=&\stackrel{2}{U}\wedge\stackrel{2}{V}+
\star(\stackrel{2}{U}\wedge\star\stackrel{2}{V})+\frac{1}{2}\com(\stackrel{2}{U},\stackrel{2}{V}),\\
\stackrel{2}{U}\stackrel{3}{V}&=&\star \stackrel{2}{U} \wedge  \star \stackrel{3}{V}-\star (\stackrel{2}{U} \wedge  \star \stackrel{3}{V}),\\
\stackrel{2}{U}\stackrel{4}{V}&=&\star \stackrel{2}{U} \wedge  \star \stackrel{4}{V},\\
\stackrel{3}{U}\stackrel{2}{V}&=&-\star \stackrel{3}{U} \wedge  \star \stackrel{2}{V}-\star (\star \stackrel{3}{U} \wedge  \stackrel{2}{V}),\\
\stackrel{3}{U}\stackrel{3}{V}&=&\star \stackrel{3}{U} \wedge  \star \stackrel{3}{V}+\star (\stackrel{3}{U} \wedge  \star \stackrel{3}{V}),\\
\stackrel{3}{U}\stackrel{4}{V}&=&\star \stackrel{3}{U} \wedge  \star \stackrel{4}{V},\\
\stackrel{4}{U}\stackrel{2}{V}&=&\star \stackrel{4}{U} \wedge  \star \stackrel{2}{V},\\
\stackrel{4}{U}\stackrel{3}{V}&=&-\star \stackrel{4}{U} \wedge  \star \stackrel{3}{V},\\
\stackrel{4}{U}\stackrel{4}{V}&=&-\star \stackrel{4}{U} \wedge  \star \stackrel{4}{V},
\end{eqnarray*}
where ranks of differential forms are denoted as $\stackrel{k}{U}\in\s{k}^\K$ and
$k=0,1,2,3,4$.
From this definition we may obtain properties of the Clifford
product of differential forms.
\begin{itemize}
\item If $U,V\in\Lambda^\K$, then $UV\in\Lambda^\K$.
\item The axioms of associativity and distributivity are satisfied
for the Clifford product.
\item $dx^\mu dx^\nu=dx^\mu\wedge dx^\nu+g^{\mu\nu},\quad
dx^\mu dx^\nu+dx^\nu dx^\mu=2 g^{\mu\nu}$.
\item $\com(U,V)=UV-VU$ for $U,V\in\s{2}^\K$.
\end{itemize}

\num{7}Let us define the trace of  differential forms as the linear operation
$\tr\,:\,\lak\to\s{0}^\K$ such that
$$
\tr(1)=1,\quad
\tr(dx^{\mu_1}\ww dx^{\mu_k})=0\quad\hbox{for}\quad k=1,2,3,4.
$$
The reader can easily prove
that
$$
\tr(UV-VU)=0,\quad \tr(V^{-1}UV)=\tr\,U,\quad U,V\in\la^\K.
$$
In the second relation $V$ is an invertible exterior form with respect
to (w.r.t.) Clifford product.

\num{8}Let us define the involution $*\,:\,\s{k}^\K\to\s{k}^\K$. By definition,
put
$$
U^*=(-1)^{\frac{k(k-1)}{2}}\check U,\quad U\in\s{k}^\K,
$$
where $\check U$ is the differential form with complex conjugated
components (if $\K=\R$, then $\check U=U$).
It is readily seen that
$$
U^{**}=U,\quad (UV)^*=V^* U^*,\quad U,V\in\lak.
$$

\num{9}Now we can define the spinor group
$$
\Spin(\V) = \{S\in \s \even\,:\,S^*S=1\}.
$$
Note that this definition is valid only for space dimensions $n<6$ (see \cite{Marchuk:Cimento}).
\num{10}Let
$$
u^{\lambda_1\ldots \lambda_r}_{\mu_1\ldots \mu_k\nu_1\ldots \nu_s }(x)
=u^{\lambda_1\ldots \lambda_r}_{[\mu_1\ldots \mu_k]\nu_1\ldots \nu_s}(x),\quad x\in\V
$$
be components of a tensor field of rank
$(r,s+k)$ antisymmetric with respect to the first
$k$ covariant indices. One may consider the following objects:
\begin{equation}
U^{\lambda_1\ldots \lambda_r}_{\nu_1\ldots \nu_s}=
\kfac u^{\lambda_1\ldots \lambda_r}_{\mu_1\ldots \mu_k\nu_1\ldots\nu_s }\,
dx^{\mu_1}\wedge\ldots\wedge dx^{\mu_k}
\label{index:form}
\end{equation}
which are formally written as
$k$-forms. Under a change of coordinates
$(x)\to(\tilde x)$ the values
(\ref{index:form}) transform as components of tensor field of
rank $(r,s)$
\begin{equation}
\tilde U^{\alpha_1\ldots \alpha_r}_{\beta_1\ldots \beta_s}=
q^{\nu_1}_{\beta_1}\ldots q^{\nu_s}_{\beta_s} p^{\alpha_1}_{\lambda_1}\ldots p^{\alpha_r}_{\lambda_r}
U^{\lambda_1\ldots \lambda_r}_{\nu_1\ldots
\nu_s},\quad q^\nu_\beta=\frac{\partial x^\nu}{\partial\tilde x^\beta},
\quad p^\alpha_\lambda=\frac{\partial\tilde x^\alpha}{\partial x^\lambda}.
\label{index:form1}
\end{equation}
The objects (\ref{index:form}) are called tensors of rank $(r,s)$ with
values in $\s{k}^\K$. We write this as
$$
U^{\lambda_1\ldots \lambda_r}_{\nu_1\ldots \nu_s}\in\s{k}^\K\top^r_s.
$$
Elements of $ \s{0}^\K\top^r_s$ are ordinary tensor fields of rank $(r,s)$ on
$\V$.
Let us define the Clifford product of the elements
$U^{\mu_1\ldots \mu_r}_{\nu_1\ldots \nu_s}\in\lak\top^r_s$ and
$V^{\alpha_1\ldots \alpha_p}_{\beta_1\ldots \beta_q}\in\lak\top^p_q$ as the
tensor field from $\lak\top^{r+p}_{s+q}$ of the form
$$
W^{\mu_1\ldots \mu_r\alpha_1\ldots \alpha_p}_{\nu_1\ldots
\nu_s\beta_1\ldots \beta_q}=
U^{\mu_1\ldots \mu_r}_{\nu_1\ldots \nu_s}
V^{\alpha_1\ldots \alpha_p}_{\beta_1\ldots \beta_q},
$$
where at right hand side there is the Clifford product of
differential forms (the indices
$\mu_1,\ldots, \mu_r,\alpha_1,\ldots, \alpha_p,\nu_1,\ldots,
\nu_s,\beta_1,\ldots, \beta_q$ are fixed).
In particular, it follows from this definition that if
$U^{\mu_1\ldots \mu_r}_{\nu_1\ldots \nu_s}\in\s{0}^\K\top^r_s$ and
$V^{\alpha_1\ldots \alpha_p}_{\beta_1\ldots \beta_q}\in\s{0}^\K\top^p_q$,
then the Clifford product of these elements is identified with
the tensor product.

Note that $dx^\mu=\delta^\mu_\nu dx^\nu$, where $\delta^\mu_\nu$ is the Kronecker
tensor ($\delta^\mu_\nu=0$ for $\mu\neq \nu$ and $\delta^\mu_\mu=1$). Hence,
$dx^\mu\in\s{1}\top^1$.

\num{11}Let us define {\sl the Upsilon derivatives} $\Upsilon_\mu$
which act on tensors from $\la^\K\top^r_s$ by the following rules:
\medskip

\noindent a) If $t_{\nu_1\ldots \nu_s}^{\epsilon_1\ldots \epsilon_r}
\in\s{0}^\K\top^r_s$, then
$$
\Upsilon_\mu t_{\nu_1\ldots \nu_s}^{\epsilon_1\ldots \epsilon_r}=
\partial_\mu t_{\nu_1\ldots \nu_s}^{\epsilon_1\ldots \epsilon_r}.
$$

\noindent b) $\Upsilon_\mu dx^\nu = -{\Gamma^\nu}_{\mu\lambda}
dx^\lambda$,
where ${\Gamma^\mu}_{\mu\lambda}={\Gamma^\mu}_{\lambda\mu}$ are Christoffel symbols (Levi-Civita
connectedness components).
\smallskip

\noindent c)  If $U^{\mu_1\ldots \mu_r}_{\nu_1\ldots
\nu_s}\in\la^\K\top^r_s$,
$V^{\alpha_1\ldots \alpha_p}_{\beta_1\ldots \beta_q}\in\la^\K\top^p_q$, then
$$
\Upsilon_\lambda(U^{\mu_1\ldots \mu_r}_{\nu_1\ldots \nu_s}
V^{\alpha_1\ldots \alpha_p}_{\beta_1\ldots \beta_q})=
(\Upsilon_\lambda U^{\mu_1\ldots \mu_r}_{\nu_1\ldots \nu_s})
V^{\alpha_1\ldots \alpha_p}_{\beta_1\ldots \beta_q}+
U^{\mu_1\ldots \mu_r}_{\nu_1\ldots \nu_s}
(\Upsilon_\lambda V^{\alpha_1\ldots \alpha_p}_{\beta_1\ldots \beta_q}).
$$
\smallskip

\noindent d)  If $U^{\mu_1\ldots \mu_r}_{\nu_1\ldots
\nu_s},
V^{\alpha_1\ldots \alpha_r}_{\beta_1\ldots \beta_s}\in\la^\K\top^r_s$, then
$$
\Upsilon_\lambda(U^{\mu_1\ldots \mu_r}_{\nu_1\ldots \nu_s}+
V^{\alpha_1\ldots \alpha_r}_{\beta_1\ldots \beta_s})=
\Upsilon_\lambda U^{\mu_1\ldots \mu_r}_{\nu_1\ldots \nu_s}+
\Upsilon_\lambda V^{\alpha_1\ldots \alpha_r}_{\beta_1\ldots \beta_s}.
$$
\medskip

With the aid of these rules it is easy to calculate how operators
$\Upsilon_\mu$ act on arbitrary tensor from
$\la^\K\top_s^r$.

From the formula $\Upsilon_\mu dx^\lambda=-{\Gamma^\lambda}_{\mu\nu}dx^\nu$ we get
\begin{equation}
(\Upsilon_\mu\Upsilon_\nu-\Upsilon_\nu\Upsilon_\mu)dx^\lambda=
-{R^\lambda}_{\rho\mu\nu}dx^\rho,
\label{R1}
\end{equation}
where
\begin{equation}
{R^\kappa}_{\lambda\mu\nu}=
\partial_\mu {\Gamma^\kappa}_{\nu\lambda}-\partial_\nu
{\Gamma^\kappa}_{\mu\lambda}+
{\Gamma^\kappa}_{\mu\eta}{\Gamma^\eta}_{\nu\lambda}-
{\Gamma^\kappa}_{\nu\eta}{\Gamma^\eta}_{\mu\lambda}
\label{R2}
\end{equation}
is rank (1,3) tensor, known as the curvature tensor (or Riemannian
tensor).

\num{12}Consider the antisymmetric tensor from $\s 2\top_2$
$$
C_{\mu\nu}=\frac{1}{2}R_{\alpha\beta\mu\nu}dx^\alpha\wedge dx^\beta,
$$
where
$R_{\alpha\beta\mu\nu}=g_{\alpha\lambda}{R^\lambda}_{\beta\mu\nu}$.
Let $B_\mu\in\s 2\top_1$ be such that
\begin{equation}
\Upsilon_\mu B_\nu-\Upsilon_\nu
B_\mu-[B_\mu,B_\nu]=\frac{1}{2}C_{\mu\nu}.
\label{B}
\end{equation}
Existence of $B_\mu$ is considered in the Theorem 2.

\num{13}Let us define the linear differential operators
$$
\D_\mu=\Upsilon_\mu-[B_\mu,\cdot\,].
$$

\theorem 1. The operators $\D_\mu$ satisfy Leibniz rule
$$
\D_\mu(UV)=(\D_\mu U)V+U\D_\mu V\quad\hbox{for}\quad U\in\la^\K\top^r_s,
\quad V\in\la^\K\top^p_q
$$
and
$$
\D_\mu \D_\nu-\D_\nu \D_\mu=0.
$$
\par

\proof\, is by direct calculation\fin

Note that the volume form $\vol $ is constant w.r.t. these operators $\D_\mu$:
$\D_\mu \vol =0$, $\mu=0,1,2,3$.

\num{14} Let differential forms $H\in\s{1}$ and
$I,K\in\s{2}$ be such that
\begin{eqnarray}
&&\D_\mu H=0,\quad\D_\mu I=0,\quad\D_\mu K=0,\quad\mu=0,1,2,3;\label{HIK:cond}\\
&&H^2=1,\, I^2=K^2=-1,\,[H,I]=[H,K]=0,\,\{I,K\}=0,\nonumber
\end{eqnarray}
where $\{I,K\}=IK+KI$.
\bigskip

We say that $dx^\mu$ are {\sl primary generators} of $\la$ and $H,\vol ,I,K$ are
{\sl secondary generators} of $\la$.

The following differential forms from
$\la$ are
linear independent at every point $x\in\V$:
\begin{equation}
H,I,K,HI,HK,IK,HIK,\ell ,\ell H,\ell I,\ell K,\ell HI,\ell HK,
\ell IK,\ell HIK,1.
\label{second:basis}
\end{equation}
These forms can be considered as basis forms of $\la$.

In Minkowski space the concept of secondary generators of Clifford algebra
$\cl(1,3)$ was presented in \cite{Marchuk:Cimento1}.

\theorem 2. On the pseudo-Riemannian space $\V$ under consideration there exists
the solution
$H\in\la_1,\,I,K\in \la_2,\,B_\mu\in\la_2\top_1$ of the system of equation
\begin{eqnarray}
&&\D_\mu B_\nu-\D_\nu B_\mu+[B_\mu,B_\nu]=\frac{1}{2}C_{\mu\nu},\label{B:eqn}\\
&&\D_\mu H=0,\quad \D_\mu I=\D_\mu K=0,\label{DHIK}\\
&&H^2=1,\,I^2=K^2=-1,\,[H,I]=[H,K]=0,\{I,\K\}=0.\label{HIK}
\end{eqnarray}

\proof. Consider local coordinates $x^\mu$ such that
\begin{equation}
g^{11}>0,\quad
\det\left|\begin{array}{cc}g^{11} &
g^{12}\\g^{12}&g^{22}\end{array}\right|<0,\quad
\det\left|\begin{array}{ccc}g^{11} & g^{12}&g^{13}\\g^{12} & g^{22}&g^{23}\\
g^{13}&g^{23}&g^{33}\end{array}\right|>0.
\label{g:condition}
\end{equation}
Let us take

\medskip

$
\noindent H = dx^{1}/\sqrt{g^{11}},
$

\begin{equation}
\displaystyle I =\frac{-( -dx^2\wedge dx^3 \,g^{11} - dx^1\wedge dx^3 \,g^{12} +
       dx^1\wedge dx^2 \,g^{13})}{
\sqrt{g^{11}}\,
       \sqrt{-((g^{13})^2\,g^{22}) + 2\,g^{12}\,g^{13}\,
          g^{23} - g^{11}\,(g^{23})^2 - (g^{12})^2\,
          g^{33} + g^{11}\,g^{22}\,g^{33}}}
\label{HIK:formulas}
\end{equation}

$K = -((dx^3\wedge dx^4 \,(g^{12})^2 - dx^2\wedge dx^4 \,g^{12}\,g^{13} +
       dx^2\wedge dx^3 \,g^{12}\,g^{14} - dx^3\wedge dx^4 \,g^{11}\,g^{22} +
dx^1\wedge dx^4 \,g^{13}\,g^{22} -
       dx^1\wedge dx^3 \,g^{14}\,g^{22} + dx^2\wedge dx^4 \,g^{11}\,
        g^{23} - dx^1\wedge dx^4 \,g^{12}\,g^{23} +
 dx^1\wedge dx^2 \,g^{14}\,g^{23} - dx^2\wedge dx^3 \,g^{11}\,
        g^{24} + dx^1\wedge dx^3 \,g^{12}\,g^{24} -
       dx^1\wedge dx^2 \,g^{13}\,g^{24})(\sqrt{-g}\,))/
      \sqrt{(g^{12})^2 - g^{11}\,g^{22}}$

\medskip

\noindent which satisfy the equalities (\ref{HIK}). Then we may consider the equalities
(\ref{DHIK}) as the linear system of equations for the components
$b_{\alpha\beta\mu}=b_{[\alpha\beta]\mu}$ of
$B_\mu=\frac{1}{2}b_{\alpha\beta\mu}dx^\alpha\wedge dx^\beta$. The number
of linear independent equations in this system of equations is equal to 24 and equal to the
number of components $b_{\alpha\beta\mu},\,\alpha<\beta$. A unique solution of this system of
equations is written in Addendum. It can be checked that this solution also satisfies
(\ref{B:eqn}). This completes the proof.
\bigskip

Let us emphasize that $H=h_\mu dx^\mu$ has the form $H=dx^1/\sqrt{g^{11}}$ only in the fixed
coordinates $x^\mu$. We suppose that under a change of coordinates the set of values
$h_\mu$ transforms as covector, i.e., $H\in\Lambda_1$. This remark is also true w.r.t. values
$I,K\in\Lambda_2,\,B_\mu\in\Lambda_2\top_1$ from (\ref{HIK:formulas}) and from Addendum.

Note that if $\{H,I,K,B_\mu\}$
is a solution of the equations (\ref{B:eqn}-\ref{HIK})
and $S\in\Spin(\V)$,
then $\{S^*HS,S^*IS,S^*KS,S^*B_\mu S-S^*\Upsilon_\mu S\}$ is also the
solution of (\ref{B:eqn}-\ref{HIK}) (see Theorem 6).

\num{15} Let us define the operation of Hermitian conjugation of differential forms
$$
U^\dagger=HU^*H,\quad U\in\la^\K.
$$
Evidently,
$$
U^{\dagger\dagger}=U,\quad (UV)^\dagger=V^\dagger U^\dagger
$$
and $\alpha^\dagger=\bar\alpha$ for $\alpha\in\s{0}^\K$, where
$\bar\alpha$ is the
complex conjugated
scalar function.

It is shown in \ref{Marchuk:Cimento1} that in Minkowski space the operation of Hermitian
conjugation of exterior forms closely connected to the operation of Hermitian conjugation of
matrices.

\num{16} Now we may define the operation
$(\,\cdot\,,\,\cdot\,)\,:\,\la^\K\times\la^\K\to\s{0}^\K$ by the formula
$$
(U,V) = \tr(UV^\dagger),\quad U,V\in\la^\K.
$$
This operation has all the properties of Hermitian scalar product at every point
$x\in\V$
\begin{eqnarray*}
&&(\alpha U,V)=\alpha(U,V),\quad (U,\alpha V)=\bar{\alpha}(U,V),\quad
(U,V)=\overline{(V,U)},\\
&& (U+W,V)=(U,V)+(W,V),\quad (U,U)>0\,\,\,\hbox{for}\,\,\,U\neq0,
\end{eqnarray*}
where $U,V,W\in\la^\K$, $\alpha\in\s{0}^\K$ and the bar means complex conjugation.

\num{17} Let us summarize properties of the operators $\D_\mu$
\begin{itemize}
\item $\D_\mu \D_\nu-\D_\nu \D_\mu=0$;
\item $\D_\mu(UV)=(\D_\mu U)V+U\D_\mu V$;
\item $\D_\mu(U+V)=\D_\mu U+\D_\mu V$;
\item $\D_\mu\ell =0$;
\item $\D_\mu H=0,\,\D_\mu I=0,\,\D_\mu K=0$;
\item $\D_\mu(U^*)=(\D_\mu U)^*$;
\item $\D_\mu(U^\dagger)=(\D_\mu U)^\dagger$;
\item $\D_\mu(\star U)=\star(\D_\mu U)$;
\item $\D_\mu(\tr U)=\tr(\D_\mu U)$;
\item $\D_\mu(U,V)=(\D_\mu U,V)+(U,\D_\mu V)$.
\end{itemize}

\num{18} In what follows we use the four spaces of differential forms
$\la^\C,\la,\s{\even}^\C,\s{\even}$. Each of these spaces can be considered at every point
$x\in\V$ as a
bialgebra with two products, namely the exterior product and Clifford product.
We use notation $\Omega$ for any of these spaces.
The operation $(\,\cdot\,,\,\cdot\,)$ converts a space $\Omega$ into
the unitary space at every point $x\in\V$.

\num{19} We say that the differential form $U\in\Omega$ is {\sl Hermitian}
if $U^\dagger=U$ and {\sl anti-Hermitian} if $U^\dagger=-U$.
Every differential form $U\in\Omega$ can be decomposed into Hermitian
and anti-Hermitian parts
$$
U=\frac{1}{2}(U+U^\dagger) +\frac{1}{2}(U-U^\dagger).
$$
Let us denote
$$
\Omega_{-} = \{U\in\Omega\,:\,U^\dagger=-U\}.
$$
This set of differential forms is closed w.r.t. commutator
$[\,\cdot\,,\,\cdot\,]$ and can be considered as a real Lie algebra.

\num{20} Let us introduce two differential forms $N,E$
\begin{eqnarray}
N&=&i\quad\hbox{for}\quad\Omega=\la^\C,\s{\even}^\C;\nonumber\\
N&=&\vol \quad\hbox{for}\quad\Omega=\la;\nonumber\\
N&=&I\quad\hbox{for}\quad\Omega=\s{\even};\label{EN:def}\\
E&=&1\quad\hbox{for}\quad\Omega=\la^\C,\la;\nonumber\\
E&=&H\quad\hbox{for}\quad\Omega=\s{\even}^\C,\s{\even}\nonumber
\end{eqnarray}
with the following properties:
\begin{eqnarray}
&&\D_\mu N=0,\quad\D_\mu E=0, \quad \mu=0,1,2,3;\nonumber\\
&&N^2=-1,\quad E^2=1,\quad [N,E]=0;\label{NE:cond}\\
&&N^\dagger=-N,\quad E^\dagger=E.\nonumber
\end{eqnarray}

\num{21} Let us define the set of differential forms
$$
L_{max}=\{U\in\Omega_-\,:\,[U,N]=[U,E]=0\}.
$$
This set of differential forms is closed w.r.t. the commutator
$[\,\cdot\,,\,\cdot\,]$
and can be considered as Lie algebra.

Let $t_1,\ldots t_p$ be generators of $L_{max}$ such that
\begin{eqnarray}
&&\D_\mu t_k=0,\quad t_k^\dagger=-t_k,\quad (t_k,t_l)=\delta_{kl},\label{tk:cond}\\
&&[t_k,N]=[t_k,E]=0,\quad
[t_k,t_l]=\sum_{q=1}^p c^q_{kl}t_q \quad k,l=1,\ldots p,\nonumber
\end{eqnarray}
where
$c^q_{kl}\,$ are real structure constants of the Lie algebra $L_{max}$.
Hence
$$
L_{max}=L(t_1,\ldots t_p)=\{\sum_{k=1}^p f_k t_k\},
$$
where $f_k=f_k(x),\,x\in\V$ are smooth real valued functions from
$C^\infty(\V,\R)$.

\num{22} Let $L_0$ be a subalgebra of the Lie algebra $L_{max}$
such that the first $d$\,\, ($d\leq p$) generators $t_1,\ldots, t_d$ of
$L_{max}$ are the generators of $L_0$
$$
L_0=L(t_1,\ldots, t_d)=\{\sum_{k=1}^n f_k t_k\}
$$
and
\begin{eqnarray*}
&& [t_k,t_l]=\sum_{q=1}^d c^q_{kl}t_q,\quad k,l=1,\ldots, d,\\
&& [t_k,t_r]=\sum_{q=d+1}^p c^q_{kr}t_q,\quad k=1,\ldots, d;\,\,
r=d+1,\ldots,p.
\end{eqnarray*}

\num{23} Let us define the set of differential forms
$$
G_0=\exp\,L_0=\{\exp\,u\,:\,u\in L_0\},
$$
where
$$
\exp\,u=1+\sum_{k=1}^\infty \frac{1}{k!}u^k.
$$
$G_0$ is a unitary ($U^{-1}=U^\dagger$) Lie group w.r.t. Clifford product
and $L_0$ is the real Lie algebra of the Lie group $G_0$.

\num{24} Let us take
$$
\alpha^\mu=H dx^\mu\in\s{\even}\top^1
$$
such that $(\alpha^\mu)^\dagger=\alpha^\mu$.


\section{Lie algebras of anti-Hermitian differential forms.}
Let $\Omega=\la^\C$. Consider the following 16 differential forms from $\Omega$:
\begin{equation}
iH,I,K,HI,HK,IK,HIK,\ell ,i\ell H,i\ell I,i\ell K,\ell HI,\ell HK,
i\ell IK,\ell HIK,i
\label{anti-Hermitian:basis}
\end{equation}
and denote them by $T_1,\ldots,T_{16}$. It is not hard to check that
\begin{equation}
\D_\mu T_k=0,\quad T_k^\dagger=-T_k,\quad (T_k,T_l)=\delta_{kl},\quad
[T_k,T_l]=c_{kl}^q T_q,
\label{T:prop}
\end{equation}
where $c_{kl}^q\,$ are real
structure constants.

Every element $U\in\Omega_-$ can be written in the form
$$
U=\sum_{k=1}^{16}(U,T_k)T_k.
$$
Hence
these differential forms $T_k$ are generators of the real Lie algebra
$\Omega_-$
$$
\Omega_-=L(T_1,\ldots T_{16})=\{\sum_{k=1}^{16}f_k T_k\},
$$
where $f_k=f_k(x),\,x\in\V$ are real valued functions from $C^\infty(\V,\R)$;
It can be shown \cite{Marchuk:AACA} that the Lie algebra
$\Omega_-$ (considered at every  point $x\in\V$) is isomorphic to the
Lie algebra $u(4)\simeq u(1)\oplus su(4)$ of anti-Hermitian $4\!\times\!4$-matrices.
So
$$
\Omega_-=L(T_{16})\oplus L(T_1,\ldots,T_{15}),
$$
where $L(T_{16})$ is isomorphic to $u(1)$ and $L(T_1,\ldots,T_{15})$ is isomorphic
to $su(4)$ (at every point $x\in\V$).

For $\Omega=\la^\C$ we have $N=i,\,E=1$. Consequently, $L_{max}=\Omega_-$.
Continuing this line of reasoning, we find real Lie algebras $\Omega_-$ and
$L_{max}$ for the cases $\Omega=\la,\s{\even}^\C,\s{\even}$.

Let us summarize our consideration of real Lie algebras of differential forms.
\begin{itemize}
\item If $\Omega=\la^\C$, then $N=i,\,E=1$ and
$$
\Omega_-=L_{max}=L(i)\oplus L(iH,I,K,HI,\ldots,\ell HIK)\simeq u(1)\oplus su(4).
$$
\item If $\Omega=\la$, then $N=\ell ,\,E=1$ and
\begin{eqnarray*}
\Omega_-&=&L(I,K,HI,HK,IK,HIK,\ell ,\ell HI,\ell HK,\ell HIK)\simeq
sp(2),\\
L_{max}&=&L(\ell )\oplus L(I,K,IK)\simeq u(1)\oplus su(2),
\end{eqnarray*}
where $sp(2)$ is the Lie algebra of the simplectic unitary Lie group.
\item If $\Omega=\s{\even}^\C$, then $N=i,\,E=H$ and
\begin{eqnarray*}
\Omega_-&=&L(i)\oplus L(\ell )\oplus
L(I,K,IK,i\ell I,i\ell K,i\ell IK)\\
&&\simeq u(1)\oplus u(1)\oplus su(2)\oplus su(2),\\
L_{max}&=&L(i)\oplus L(I,K,IK)\simeq u(1)\oplus su(2),
\end{eqnarray*}
\item If $\Omega=\s{\even}$, then $N=I,\,E=H$ and
\begin{eqnarray*}
\Omega_-&=&L(I)\oplus L(\ell )\simeq u(1)\oplus u(1),\\
L_{max}&=&L(I)\simeq u(1),
\end{eqnarray*}
\end{itemize}

Let us consider differential forms $t_1,\ldots,t_{16}$ such
that $t_k=\sum_{p=1}^{16}r^p_{k} T_p$, where real constants $r^p_k$ are
elements of an orthogonal $16\!\times\!16$-matrix. Evidently
$$
\D_\mu t_k=0,\quad t_k^\dagger=-t_k,\quad (t_k,t_l)=\delta_{kl}
$$
and the set of differential forms $\{t_1,\ldots,t_{16}\}$ is another
set of generators of the real Lie algebra $\Omega_-\simeq u(4)$ in case
$\Omega=\la^\C$. We claim that this construction allow us to describe a
Lie subalgebra of $\Omega_-$ that is isomorphic to the Lie algebra
$su(3)$. Indeed, let us take the following generators of $\Omega_-$:
\begin{eqnarray}
t_{1} &=& (H I K + I K)/\sqrt{2},\nonumber\\
t_{2} &=& (H K + K)/\sqrt{2},\nonumber\\
t_{3} &=& (-H I - I)/\sqrt{2},\nonumber\\
t_{4} &=& (\ell  - \ell  I i)/\sqrt{2},\nonumber\\
t_{5} &=& (\ell  H I + \ell  H i)/\sqrt{2},\nonumber\\
t_{6} &=& (-\ell  H K + \ell  I K i)/\sqrt{2},\nonumber\\
t_{7} &=& (-\ell  H I K - \ell  K i)/\sqrt{2},\nonumber\\
t_{8} &=& (-H I - 2 H i + I)/\sqrt{6},\label{su(3):generators}\\
t_{9} &=& (\ell  H K + \ell  I K i)/\sqrt{2},\nonumber\\
t_{10} &=& (-\ell  H I K + \ell  K i)/\sqrt{2},\nonumber\\
t_{11} &=& (\ell  + \ell  I i)/\sqrt{2},\nonumber\\
t_{12} &=& (-\ell  H I + \ell  H i)/\sqrt{2},\nonumber\\
t_{13} &=& (-H I K + I K)/\sqrt{2},\nonumber\\
t_{14} &=& (-H K + K)/\sqrt{2},\nonumber\\
t_{15} &=& (H I - H i - I)/\sqrt{3},\nonumber\\
t_{16} &=& i\nonumber,
\end{eqnarray}
which are expressed via generators $T_1,\ldots,T_{16}$ from
(\ref{anti-Hermitian:basis})
with the aid of the orthogonal matrix. It can be easily checked that
differential forms $t_1,\ldots,t_8$ are the  generators of the real Lie
algebra $L_0 =L(t_1,\ldots,t_8)$ isomorphic to the Lie algebra
$su(3)$ and
\begin{eqnarray*}
&& [t_k,t_l] =\sum_{q=1}^8 c^q_{kl}t_q,\quad k,l=1,\ldots, 8,\\
&& [t_k,t_r] =\sum_{q=9}^{15} c^q_{kr}t_q,\quad k=1,\ldots, 8;\,\,
r=9,\ldots,16.
\end{eqnarray*}


\section{Main equations.}
Now we may write down the main system of equations
\begin{equation}
\D_\mu B_\nu-\D_\nu B_\mu +[B_\mu,B_\nu]=\frac{1}{2}C_{\mu\nu},
\label{B:def}
\end{equation}

\begin{eqnarray}
&&\D_\mu N=0,\quad\D_\mu E=0, \quad \mu=0,1,2,3;\label{NE}\\
&&N^2=-1,\quad E^2=1,\quad [N,E]=0; \quad N^\dagger=-N,\quad E^\dagger=E.\label{NT1}
\end{eqnarray}

\begin{equation}
dx^\mu(\D_\mu\Psi+\Psi A_\mu+B_\mu\Psi)N-m\Psi E=0,\label{Dirac:eq}
\end{equation}

\begin{eqnarray}
&&\D_\mu A_\nu-\D_\nu A_\mu-[A_\mu,A_\nu]=F_{\mu\nu},\label{YM1:eq}\\
&&\frac{1}{\sqrt{-g}}\D_\mu(\sqrt{-g}F^{\mu\nu})-[A_\mu,F^{\mu\nu}]=J^\nu,
\label{YM2:eq}
\end{eqnarray}

\begin{equation}
J^\nu=\frac{1}{4}\sum_{k=1}^d(E\{E,\{N,\Psi^\dagger\alpha^\nu\Psi\}\},t_k)t_k,
\label{J:def}
\end{equation}

where $\Psi\in\Omega$, $A_\mu\in L_0\top_1$, $F_{\mu\nu}\in
L_0\top_2$,
$F^{\mu\nu}=g^{\mu\alpha}g^{\nu\beta}F_{\alpha\beta}\in
L_0\top^2$, $B_\mu\in\s{2}\top_1$, $m$ is a real
constant. The values $J^\nu=J^\nu(\Psi,E,N)\in L_0\top^1$ also can be
defined with the aid of the following equalities:
\begin{eqnarray}
J^\nu_{(1)}&=&\Psi^\dagger\alpha^\nu\Psi\in\Omega\top^1\,\,:\,\,
(J^\nu_{(1)})^\dagger=J^\nu_{(1)}.\nonumber\\
J^\nu_{(2)}&=&\frac{1}{2}\{N,J^\nu_{(1)}\}\in\Omega_-\top^1\,\,:\,\,
[J^\nu_{(2)},N]=0,\label{J:def1}\\
J^\nu_{(3)}&=&\frac{1}{2}E\{E,J^\nu_{(2)}\}\in L_{max}\top^1
\,\,:\,\,[J^\nu_{(3)},N]=[J^\nu_{(3)},E]=0,\nonumber\\
J^\nu&=&\sum_{k=1}^d(J^\nu_{(3)},t_k)t_k\in L_0\top^1,\nonumber
\end{eqnarray}
where $t_1,\ldots t_d$ are the generators (\ref{tk:cond})
of the real Lie algebra
$L_0$ (see subsection 22 of section 1).

We say that the equation (\ref{Dirac:eq}) is {\sl a Dirac type tensor
equation} and the equations (\ref{YM1:eq},\ref{YM2:eq}) are
{\sl Yang-Mills equations}.

The full set of $\{A_\mu,F_{\mu\nu}\}$ is called {\sl the Yang-Mills
field}; $A_\mu$ is potential and $F_{\mu\nu}$ is strength of the
Yang-Mills field.

In the Standard Model of elementary particles three real Lie algebras
are of special interest. Namely the Lie algebra $su(3)$ is used in Quantum
Chromodynamics, $u(1)\oplus su(2)$ is used in Electroweak Theory, and
$u(1)$ is used in Quantum Electrodynamics. Taking these Lie algebras
into account, we consider five special cases of equations
(\ref{B:def}-\ref{J:def}).
\begin{description}
\item{(i)} $\Omega=\la^\C,\, N=i,\,E=1$,
$$
L_0=L_{max}\simeq u(1)\oplus su(4),\quad
J^\nu=i\Psi^\dagger\alpha^\nu\Psi;
$$
\item{(ii)} $\Omega=\la^\C,\, N=i,\,E=1$,
$$
L_0=L(t_1,\ldots,t_8)\simeq su(3),\quad
J^\nu=\sum_{k=1}^8(i\Psi\alpha^\nu\Psi,t_k)t_k,
$$
where $t_1,\ldots,t_8$ from (\ref{su(3):generators});
\item{(iii)} $\Omega=\s{\even}^\C,\,N=i,\,E=H$,
$$
L_0=L_{max}\simeq u(1)\oplus su(2),\quad
J^\nu=\frac{1}{2}H\{H,i\Psi^\dagger\alpha^\nu\Psi\};
$$
\item{(iv)} $\Omega=\la,\,N=\ell ,\,E=1$,
$$
L_0=L_{max}\simeq u(1)\oplus su(2),\quad
J^\nu=\frac{1}{2}\{\ell ,\Psi^\dagger\alpha^\nu\Psi\};
$$
\item{(v)} $\Omega=\s{\even},\,N=I,\,E=H$,
$$
L_0=L_{max}=L(I)\simeq u(1),\quad
J^\nu=\frac{1}{4}H\{H,\{I,\Psi^\dagger\alpha^\nu\Psi\}\}.
$$
\end{description}

The full set of $\{\Psi,E,N\}$ is called
{\sl the wave field} of the equations (\ref{NE}-\ref{Dirac:eq}).

In the system of equations  (\ref{B:def}-\ref{J:def}) we consider
the wave field $\{\Psi,E,N\}$, the Yang-Mills
field $\{A_\mu,F_{\mu\nu}\}$, and the field $B_\mu$
as unknown values.


\section{Nonabelian charge conservation laws.}
\theorem 3. If the values $\Psi\in\Omega$, $A_\mu\in L_0\top_1$,
$B_\mu\in\s{2}\top_1$, $N\in\Omega$, $E\in\Omega$ satisfy Dirac type
tensor equation (\ref{Dirac:eq}) and the values $J^\mu\in L_0\top^1$
are defined
in (\ref{J:def}) or, equivalently, in (\ref{J:def1}),
then $J^\mu$ satisfy the equality
\begin{equation}
\frac{1}{\sqrt{-g}}\D_\mu(\sqrt{-g}J^\mu)-[A_\mu,J^\mu]=0.
\label{conserv:law}
\end{equation}
\par

The equality (\ref{conserv:law}) is called {\sl the nonabelian charge
conservation law} for the Dirac type tensor equation.
\medskip

\proof. Let us multiply the equation (\ref{Dirac:eq}) from the left by $H$
and from the right by $-N$. We denote the left hand side of resulting equation  by
\begin{equation}
Q = \alpha^\mu(\D_\mu\Psi+\Psi A_\mu+B_\mu \Psi)+m H\Psi EN.
\label{Q:def}
\end{equation}
Then
$$
Q^\dagger=(\D_\mu\Psi^\dagger-A_\mu\Psi^\dagger+\Psi^\dagger
B_\mu^\dagger)\alpha^\mu-mNE\Psi^\dagger H.
$$
Consider the expression
\begin{eqnarray*}
Y_{(1)} &=& \Psi^\dagger Q+Q^\dagger\Psi\\
&=&\D_\mu(\Psi^\dagger\alpha^\mu\Psi)+\Psi^\dagger(-\D_\mu\alpha^\mu
+\alpha^\mu B_\mu+B_\mu^\dagger\alpha^\mu)\Psi-
[A_\mu,\Psi^\dagger\alpha^\mu\Psi]+m[\Psi^\dagger H\Psi,EN]\\
&=&\D_\mu(\Psi^\dagger\alpha^\mu\Psi)+
\Gamma^\mu_{\mu\nu}\Psi^\dagger\alpha^\nu\Psi-
[A_\mu,\Psi^\dagger\alpha^\mu\Psi]+m[\Psi^\dagger H\Psi,EN]\\
&=&\frac{1}{\sqrt{-g}}\D_\mu(\sqrt{-g}J^\mu_{(1)})-[A_\mu,J^\mu_{(1)}]
+m[\Psi^\dagger H\Psi,EN].
\end{eqnarray*}
Here we use the formulas
\begin{eqnarray*}
&&\D_\mu\alpha^\mu=-\Gamma^\mu_{\mu\nu}\alpha^\nu
+\alpha^\mu B_\mu+B_\mu^\dagger\alpha^\mu,\\
&&\D_\mu J^\mu_{(1)}+\Gamma^\mu_{\mu\nu} J^\nu_{(1)}=
\frac{1}{\sqrt{-g}}\D_\mu(\sqrt{-g}J^\mu_{(1)}),
\end{eqnarray*}
see \cite{Marchuk:Cimento}.

If we take
\begin{eqnarray*}
Y_{(2)} &=& \frac{1}{2}\{N,Y_{(1)}\},\\
Y_{(3)} &=& \frac{1}{2}E\{E,Y_{(2)}\},
\end{eqnarray*}
then we get
$$
Y_{(3)}=\frac{1}{\sqrt{-g}}\D_\mu(\sqrt{-g}J^\mu_{(3)})-[A_\mu,J^\mu_{(3)}]
+\frac{1}{4}m E\{E,\{N,[\Psi^\dagger H\Psi,EN]\}\}.
$$
Using the equalities $N^2=-1$, $E^2=1$, $[N,E]=0$, it is easy to check
that
$$
E\{E,\{N,[\Psi^\dagger H\Psi,EN]\}\}\equiv 0.
$$
Hence,
$$
Y_{(3)}=\frac{1}{\sqrt{-g}}\D_\mu(\sqrt{-g}J^\mu_{(3)})-[A_\mu,J^\mu_{(3)}].
$$

Finally, we take
$$
Y = \sum_{k=1}^d(Y_{(3)},t_k)t_k,
$$
where the generators $t_1,\ldots,t_d$ of the Lie algebra $L_0$ are
defined in  the subsection 22 of section 1.
In this case we obtain
$$
Y=\frac{1}{\sqrt{-g}}\D_\mu(\sqrt{-g}J^\mu)-[A_\mu,J^\mu].
$$

By assumption, $\Psi,N,E,A_\mu,B_\mu$ satisfy (\ref{Dirac:eq}). Consequently,
$Q=Y_{(1)}=Y_{(2)}=Y_{(3)}=Y=0$ and $J^\mu$ satisfy equality
(\ref{conserv:law}).
This completes the proof.
\bigskip

\theorem 4. Let us denote the left hand side of the equation (\ref{YM2:eq})
by $R^\nu$
$$
R^\nu
=\frac{1}{\sqrt{-g}}\D_\mu(\sqrt{-g}F^{\mu\nu})-[A_\mu,F^{\mu\nu}],
$$
where $F_{\mu\nu}$ satisfy (\ref{YM1:eq}). Then
$$
\frac{1}{\sqrt{-g}}\D_\mu(\sqrt{-g}\,R^\mu)-[A_\mu,R^\mu]=0.
$$
The proof is by direct calculation.
\bigskip

This theorem means that the equation (\ref{YM2:eq}) is consistent with the
nonabelian charge conservation law (\ref{conserv:law}).


\section{Unitary and Spin gauge symmetries.}
\theorem 5. Suppose that the following conditions hold:
\begin{enumerate}
\item The values
\begin{eqnarray}
&&\Psi\in\Omega,\,A_\mu\in L_0\top_1,\,F_{\mu\nu}\in L_0\top_2,\,J^\nu\in L_0\top^1,\nonumber\\
&&E\in\Omega,\,N\in\Omega,\,B_\mu\in\s{2}\top_1,\,C_{\mu\nu}\in\s{2}\top_2
\label{Values}
\end{eqnarray}
and operators $\D_\mu=\Upsilon_\mu - [B_\mu,\,\cdot\,]$ satisfy equations
(\ref{B:def}-\ref{J:def}).
\item $U=U(x)$ is a smooth map $\V\to G_0$, where the Lie group $G_0$ is defined in subsection 23
of section 1.
\end{enumerate}
Then the values with acute
\begin{eqnarray}
&&\acute{\Psi}=\Psi U,\,\acute{A}_\mu=U^{-1}A_\mu U-U^{-1}\D_\mu U,\,
\acute{F}_{\mu\nu}=
U^{-1}F_{\mu\nu}U,\,\acute{J}^\nu=U^{-1}J^\nu U,\nonumber\\
&&\{\acute{E},\acute{N},\acute{B}_\mu,\acute{C}_{\mu\nu}\}=
\{E,N,B_\mu,C_{\mu\nu}\}
\label{prime:Values}
\end{eqnarray}
and the operators $\acute{\D}_\mu=\D_\mu$ satisfy the same equations (\ref{B:def}-\ref{J:def}).
\par
\bigskip

The proof is straightforward.

\theorem 6. If the values (\ref{Values}) and operators $\D_\mu$ satisfy the equations
(\ref{B:def}-\ref{J:def})
and $S=S(x)$ is a smooth map
$\V\to\Spin(\V)$, then the values with check
\begin{eqnarray}
&&\check{\Psi}=\Psi S,\,\check{A}_\mu=S^{-1}A_\mu S,\,\check{F}_{\mu\nu}=
S^{-1}F_{\mu\nu}S,\,\check{J}^\nu=S^{-1}J^\nu S,\nonumber\\
&&\check{E}=S^{-1}ES,\,\check{N}=S^{-1}NS,\,
\check{B}_\mu=S^{-1}B_\mu S-S^{-1}\Upsilon_\mu S,\,
\check{C}_{\mu\nu}=C_{\mu\nu}
\label{check:Values}
\end{eqnarray}
and the operators $\check{\D}_\mu=\Upsilon_\mu-[\check{B}_\mu,\,\cdot\,]$
satisfy the same equations (\ref{B:def}-\ref{J:def}).
\par
\bigskip

The proof is by direct calculation.
Note that for values with check the operation of Hermitian conjugation is defined by
$$
\check{U}^\dagger = \check{H}\check{U}^*\check{H},
$$
where $\check{H}=S^{-1}HS$.

Invariance of the equation (\ref{B:def}) is proved in \cite{Marchuk:Cimento}.


\section{Lagrangians.}
Denote
\begin{equation}
P=dx^\mu(\D_\mu\Psi+\Psi A_\mu+B_\mu\Psi)N-m\Psi E.
\end{equation}
We have $\Psi\in\Omega$
\begin{eqnarray*}
&&\Psi=\psi+\psi_\mu dx^\mu+\frac{1}{2!}\psi_{\mu_1\mu_2}dx^{\mu_1}\wedge dx^{\mu_2} +
\ldots +\frac{1}{4!}\psi_{\mu_1\ldots\mu_4}dx^{\mu_1}\wedge\ldots\wedge dx^{\mu_4},\\
&&\psi_{\mu_1\ldots\mu_k}=\psi_{[\mu_1\ldots\mu_k]}=p_{\mu_1\ldots\mu_k}+i
q_{\mu_1\ldots\mu_k},
\end{eqnarray*}
where $p_{\mu_1\ldots\mu_k},\,\,q_{\mu_1\ldots\mu_k}$ are real valued components of a covariant
antisymmetric tensor fields of rank $k$ and
\begin{eqnarray*}
q_{\mu_1\ldots\mu_k}&\equiv& 0\quad(k=0,\ldots 4)\quad\hbox{for}\quad\Omega=\Lambda,\Lambda_\even,\\
\psi_\mu,\psi_{\mu_1\mu_2\mu_3}&\equiv&0\quad\hbox{for}\quad\Omega=\Lambda_\even^\C,\Lambda_\even.
\end{eqnarray*}
Consider the Lagrangian (Lagrangian density)
$$
\L_0=\frac{\sqrt{-g}}{4}\tr(H(\Psi^*P+P^*\Psi)).
$$
A Dirac type tensor equation can be derived from $\L_0$ with the aid of the following formulas:
\begin{eqnarray*}
&&u^{\mu_1\ldots\mu_k}=(\frac{\partial\L_0}{\partial p_{\mu_1\ldots\mu_k}}-
\partial_\nu\frac{\partial\L_0}{\partial p_{\mu_1\ldots\mu_k,\nu}})+
i(\frac{\partial\L_0}{\partial q_{\mu_1\ldots\mu_k}}-
\partial_\nu\frac{\partial\L_0}{\partial q_{\mu_1\ldots\mu_k,\nu}}),\\
&&u_{\nu_1\ldots\nu_k}=u_{[\nu_1\ldots\nu_k]}=
g_{\mu_1\nu_1}\ldots g_{\mu_k\nu_k}u^{\mu_1\ldots\mu_k},\\
&&U=\sum^4_{k=0}\frac{1}{k!}u_{\nu_1\ldots\nu_k}dx^{\nu_1}\ww dx^{\nu_k},\\
&&\acute{P}=\frac{1}{\sqrt{-g}}U H,
\end{eqnarray*}
where $p_{\mu_1\ldots\mu_k,\nu}=\partial_\nu p_{\mu_1\ldots\mu_k},\,\,
q_{\mu_1\ldots\mu_k,\nu}=\partial_\nu q_{\mu_1\ldots\mu_k}$. Now we may check that
$\acute{P}\equiv P$.

Let us consider the Lagrangian for Yang-Mills equations. Denote
\begin{eqnarray*}
F_{\mu\nu}&=&\D_\mu A_\nu-\D_\nu A_\mu-[A_\mu,A_\nu],\\
F^{\alpha\beta}&=&g^{\alpha\mu}g^{\beta\nu}F_{\mu\nu},\\
Y^\nu&=&\frac{1}{\sqrt{-g}}\D_\mu(\sqrt{-g}\,F^{\mu\nu})-[A_\mu,F^{\mu\nu}]-J^\nu,
\end{eqnarray*}
where $J^\nu$ are from (\ref{J:def}) and $A_\mu=a_\mu^k t_k$, ($t_k=t^k$ are defined in the
subsection 22 of section 1).

The Yang-Mills Lagrangian has the form
$$
\L_1=\frac{\sqrt{-g}}{4}\tr(F_{\mu\nu}F^{\mu\nu})
$$
and the complete Lagrangian for the Dirac type tensor equation together with Yang-Mills equations
$$
\L=\L_1+2n\L_0.
$$
Here $n=4$ is the dimension of space-time.
Denote
\begin{eqnarray*}
y^\mu_k&=&\frac{\partial\L}{\partial a^k_\mu}-
\partial_\nu\frac{\partial\L}{\partial a^k_{\mu,\nu}},\\
\acute{Y}^\nu&=&\frac{1}{\sqrt{-g}}y^\mu_k t^k.
\end{eqnarray*}
Now we may check that
$$
\acute{Y}^\nu\equiv Y^\nu.
$$
Hence the Dirac type tensor equation and the Yang-Mills equations can be derived from
the Lagrangian $\L$.

We hope that equations (\ref{B:def}-\ref{J:def})
can be used to describe fermions in space-time
with presence of gravitation. From this point of view it is reasonable to add
the Einstein-Hilbert Lagrangian $\sqrt{-g}\,R$ to  $\L$ and variate
the resulting Lagrangian w.r.t. components of metric tensor $g_{\mu\nu}$.


\section{A geometrical interpretation of results.}
We consider a pseudo-Riemannian space $\V$ with the Levi-Civita
connection ${\Gamma^\lambda}_{\mu\nu}$, with the covariant derivatives
$\nabla_\mu$, with the Upsilon derivatives $\Upsilon_\mu$, and with the
curvature tensor $R_{\alpha\beta\mu\nu}$ defined in section 1.
Suppose that a new structure on $\V$ is given. Namely the affine
connection ${\check{\Gamma}^\lambda}{}_{\mu\nu}$. We get definitions of
the covariant derivatives
$\check{\nabla}_\mu$, the Upsilon derivatives $\check{\Upsilon}_\mu$,
and the
curvature tensor $\check{R}_{\alpha\beta\mu\nu}$, replacing ${\Gamma^\lambda}_{\mu\nu}$
by ${\check{\Gamma}^\lambda}{}_{\mu\nu}$ in the corresponding definitions in
section 1. We suppose that the affine connection
${\check{\Gamma}^\lambda}{}_{\mu\nu}$ is metric compatible
$$
\check{\nabla}_\kappa g_{\mu\nu}=0,\quad
\check{\nabla}_\kappa g^{\mu\nu}=0.
$$
It is convenient to introduce the tensor
$$
{K^\lambda}_{\mu\nu}={\check{\Gamma}^\lambda}{}_{\mu\nu}-{\Gamma^\lambda}_{\mu\nu},
$$
which we, following \cite{Nakahara}, will call contorsion.
It is easy to see that the affine connection is metric compatible iff
$K_{\nu\mu\lambda}=-K_{\lambda\mu\nu}$. Torsion is expressed via
contorsion as
$$
{T^\lambda}_{\mu\nu}={K^\lambda}_{\mu\nu}-{K^\lambda}_{\nu\mu}.
$$
Conversely, the contorsion of a metric compatible connection is
expressed via torsion as
$$
K^\lambda{}_{\mu\nu}=\frac{1}{2}(T^\lambda{}_{\mu\nu}+T_\mu{}^\lambda{}_\nu+
T_\nu{}^\lambda{}_\mu)
$$
(see \cite{Nakahara}, formula (7.35)).

So we arrive at the affine space
$\{\M,g_{\mu\nu},{K^\lambda}_{\mu\nu}\}$.
Let us define the tensors
\begin{eqnarray*}
b_{\alpha\beta\mu}&=&-\frac{1}{2}K_{\alpha\mu\beta},\\
B_\mu &=& \frac{1}{2}b_{\alpha\beta\mu}dx^\alpha\wedge
dx^\beta\in\Lambda^2\top_1,\\
C_{\mu\nu} &=& \frac{1}{2}R_{\alpha\beta\mu\nu}dx^\alpha\wedge
dx^\beta\in\Lambda^2\top_2.
\end{eqnarray*}

\theorem 7. If $U\in\Lambda$, then
$$
\check{\Upsilon}_\mu U=\Upsilon_\mu U - [B_\mu,U].
$$
\par

\proof\,\, follows from the formula
$$
{K^\nu}_{\mu\lambda} dx^\lambda=[B_\mu,dx^\nu],
$$
which can be easily checked.
\bigskip

\theorem 8. (F.E.Burstall, A.D.King, N.G.Marchuk, D.G.Vassiliev)
The following equality holds
$$
\Upsilon_\mu B_\nu-\Upsilon_\nu
B_\mu-[B_\mu,B_\nu]=\frac{1}{2}C_{\mu\nu}
$$
iff
$$
\check{R}_{\alpha\beta\mu\nu}=0.
$$
\proof. Suppose that the tensors $q_{\alpha\beta\mu\nu}$ and
$\check{R}_{\alpha\beta\mu\nu}$ are such that
$$
\frac{1}{2}q_{\alpha\beta\mu\nu}dx^\alpha\wedge
dx^\beta=\Upsilon_\mu B_\nu-\Upsilon_\nu
B_\mu-[B_\mu,B_\nu]-\frac{1}{2}C_{\mu\nu}
$$
and
\begin{equation}
\check{R}_{\alpha\lambda\mu\nu}=g_{\kappa\alpha}(
\partial_\mu {\check{\Gamma}^\kappa}{}_{\nu\lambda}-\partial_\nu
{\check{\Gamma}^\kappa}{}_{\mu\lambda}+
{\check{\Gamma}^\kappa}{}_{\mu\eta}{\check{\Gamma}^\eta}{}_{\nu\lambda}-
{\check{\Gamma}^\kappa}{}_{\nu\eta}{\check{\Gamma}^\eta}{}_{\mu\lambda}).
\label{check:R}
\end{equation}
Then
$$
\check{R}_{\alpha\beta\mu\nu}=-2q_{\alpha\beta\mu\nu}.
$$
This completes the proof.
\bigskip


\section{Space dimensions $n=2,3$.}
Let $\M_3$ be a three dimensional differentiable manifolds with local
coordinates $x^1,x^2,x^3$ and with the smooth metric tensor
$g_{\mu\nu}=g_{\nu\mu},\,(\mu,\nu=1,2,3)$ that satisfy
(\ref{g:condition}). Denote by $\V_3$ the pseudo-Riemannian space
$\{\M_3,g\}$ with the Levi-Civita connection
$\Gamma^\lambda{}_{\mu\nu}$, with the covariant derivatives $\nabla_\mu$,
with the Upsilon derivatives $\Upsilon_\mu$, and with the curvature
tensor $R_{\alpha\beta\mu\nu}$.

\theorem 10. On the pseudo-Riemannian space $\V_3$ there exists the
solution $H\in\Lambda_1$, $I\in\Lambda_2$, $B_\mu\in\Lambda_2\top_1$ of
the system of equations that consists of (\ref{B:def}) and
\begin{eqnarray}
&&\D_\mu H=0,\quad \D_\mu I=0,\label{DHI}\\
&&H^2=1,\quad I^2=-1,\quad [H,I]=0,\label{HI:conditions}
\end{eqnarray}
where $\D_\mu=\Upsilon_\mu-[B_\mu,\,\cdot\,]$.\par

\proof. Evidently $H,I$ from (\ref{HIK:formulas}) satisfy
(\ref{HI:conditions}). It can be checked that the following components
$b_{\alpha\beta\mu}=b_{[\alpha\beta]\mu}$ of
$B_\mu=\frac{1}{2}b_{\alpha\beta\mu}dx^\alpha\wedge dx^\beta$ satisfy
(\ref{B:def}) and (\ref{DHI}):

\medskip

$b_{12 1} =  (\partial_{1}g_{33}\,g_{13}\,g_{22}\,g_{23} -
     \partial_{1}g_{33}\,g_{12}\,g_{23}^2 -
     2\,\partial_{1}g_{23}\,g_{13}\,g_{22}\,g_{33} +
     2\,\partial_{1}g_{23}\,g_{12}\,g_{23}\,g_{33} +
     \partial_{1}g_{22}\,g_{13}\,g_{23}\,g_{33} -
     \partial_{2}g_{11}\,g_{23}^2\,g_{33} -
     \partial_{1}g_{22}\,g_{12}\,g_{33}^2 +
     \partial_{2}g_{11}\,g_{22}\,g_{33}^2)/
   (4\,g_{33}\,(-g_{23}^2 +
       g_{22}\,g_{33}))$

$b_{12 2} =  (\partial_{2}g_{33}\,g_{13}\,g_{22}\,g_{23} -
     \partial_{2}g_{33}\,g_{12}\,g_{23}^2 -
     2\,\partial_{2}g_{23}\,g_{13}\,g_{22}\,g_{33} +
     2\,\partial_{2}g_{23}\,g_{12}\,g_{23}\,g_{33} +
     \partial_{2}g_{22}\,g_{13}\,g_{23}\,g_{33} +
     \partial_{1}g_{22}\,g_{23}^2\,g_{33} -
     2\,\partial_{2}g_{12}\,g_{23}^2\,g_{33} -
     \partial_{2}g_{22}\,g_{12}\,g_{33}^2 -
     \partial_{1}g_{22}\,g_{22}\,g_{33}^2 +
     2\,\partial_{2}g_{12}\,g_{22}\,g_{33}^2)/
   (4\,g_{33}\,(-g_{23}^2 +
       g_{22}\,g_{33}))$

$b_{12 3} =  (\partial_{3}g_{33}\,g_{13}\,g_{22}\,g_{23} -
     \partial_{3}g_{33}\,g_{12}\,g_{23}^2 -
     2\,\partial_{3}g_{23}\,g_{13}\,g_{22}\,g_{33} +
     2\,\partial_{3}g_{23}\,g_{12}\,g_{23}\,g_{33} +
     \partial_{3}g_{22}\,g_{13}\,g_{23}\,g_{33} +
     \partial_{1}g_{23}\,g_{23}^2\,g_{33} -
     \partial_{2}g_{13}\,g_{23}^2\,g_{33} -
     \partial_{3}g_{12}\,g_{23}^2\,g_{33} -
     \partial_{3}g_{22}\,g_{12}\,g_{33}^2 -
     \partial_{1}g_{23}\,g_{22}\,g_{33}^2 +
     \partial_{2}g_{13}\,g_{22}\,g_{33}^2 +
     \partial_{3}g_{12}\,g_{22}\,g_{33}^2)/
   (4\,g_{33}\,(-g_{23}^2 +
       g_{22}\,g_{33}))$

$b_{13 1} =  (-(\partial_{1}g_{33}\,g_{13}) + \partial_{3}g_{11}\,g_{33})/
   (4\,g_{33})$

$b_{13 2} =  (-(\partial_{2}g_{33}\,g_{13}) - \partial_{1}g_{23}\,g_{33} +
     \partial_{2}g_{13}\,g_{33} + \partial_{3}g_{12}\,g_{33})/(4\,g_{33})$

$b_{13 3} =  (-(\partial_{3}g_{33}\,g_{13}) - \partial_{1}g_{33}\,g_{33} +
     2\,\partial_{3}g_{13}\,g_{33})/(4\,g_{33})$

$b_{23 1} =  -(\partial_{1}g_{33}\,g_{23} - \partial_{1}g_{23}\,g_{33} +
      \partial_{2}g_{13}\,g_{33} - \partial_{3}g_{12}\,g_{33})/(4\,g_{33})$

$b_{23 2} =  -(\partial_{2}g_{33}\,g_{23} - \partial_{3}g_{22}\,g_{33})/
   (4\,g_{33})$

$b_{23 3} =  -(\partial_{3}g_{33}\,g_{23} + \partial_{2}g_{33}\,g_{33} -
      2\,\partial_{3}g_{23}\,g_{33})/(4\,g_{33})$

\medskip

Note that $\{S^*HS,S^*IS,S^*B_\mu S-S^*\Upsilon_\mu S\}$ also satisfy
(\ref{DHI}-\ref{HI:conditions}), where $S\in\Spin(\V_3)$.
\bigskip

Let $\M_2$ be a two dimensional differentiable manifolds with local
coordinates $x^1,x^2$ and with the smooth metric tensor
$g_{\mu\nu}=g_{\nu\mu},\,(\mu,\nu=1,2)$ such that
$g_{11}>0$, $g_{11}g_{22}-(g_{12})^2<0$.
Denote by $\V_2$ the pseudo-Riemannian space
$\{\M_2,g\}$ with
$\Gamma^\lambda{}_{\mu\nu},\nabla_\mu,
\Upsilon_\mu,\D_\mu,R_{\alpha\beta\mu\nu},C_{\mu\nu}$ defined in section
1.

\theorem 11. On the pseudo-Riemannian space $\V_2$ there exists the
solution $H\in\Lambda_1$, $B_\mu\in\Lambda_2\top_1$ of
the system of equations that consists of (\ref{B:def}) and
\begin{equation}
\D_\mu H=0,\quad
H^2=1.\label{H}
\end{equation}
\par

\proof. We may take

\begin{eqnarray*}
b_{12 1} &=&  -(\partial_{1}g_{22}\,g_{12} -
\partial_{2}g_{11}\,g_{22})/(4\,g_{22})\cr
b_{12 2} &=&  -(\partial_{2}g_{22}\,g_{12} + \partial_{1}g_{22}\,g_{22} -
      2\,\partial_{2}g_{12}\,g_{22})/(4\,g_{22})
\,)
\end{eqnarray*}
and check that $H=dx^1/\sqrt{g^{11}}$ and
$B_\mu=\frac{1}{2}b_{\alpha\beta\mu}dx^\alpha\wedge dx^\beta$ satisfy
(\ref{B:def}) and (\ref{H}).


\section{Addendum.}
Here are formulas for the components $b_{\alpha\beta\mu}$ of $B_\mu$.
These $B_\mu$ together with the $H,I,K$ from (\ref{HIK:formulas}) satisfy
equations (\ref{B:eqn}-\ref{HIK})

\medskip

$b_{12 1} = (\partial_1g_{44}\,g_{14}\,g_{23}\,g_{24}^2\,g_{33}\,g_{34} - \partial_1g_{44}\,g_{13}\,g_{24}^3\,g_{33}\,g_{34} -
      2\,\partial_1g_{44}\,g_{14}\,g_{23}^2\,g_{24}\,g_{34}^2 + 2\,\partial_1g_{44}\,g_{13}\,g_{23}\,g_{24}^2\,g_{34}^2 +
      \partial_1g_{44}\,g_{14}\,g_{22}\,g_{23}\,g_{34}^3 - \partial_1g_{44}\,g_{13}\,g_{22}\,g_{24}\,g_{34}^3 +
      \partial_1g_{44}\,g_{14}\,g_{23}^2\,g_{24}\,g_{33}\,g_{44} - \partial_1g_{44}\,g_{13}\,g_{23}\,g_{24}^2\,g_{33}\,g_{44} -
      2\,\partial_1g_{34}\,g_{14}\,g_{23}\,g_{24}^2\,g_{33}\,g_{44} + 2\,\partial_1g_{34}\,g_{13}\,g_{24}^3\,g_{33}\,g_{44} -
      \partial_1g_{44}\,g_{14}\,g_{22}\,g_{24}\,g_{33}^2\,g_{44} + \partial_1g_{44}\,g_{12}\,g_{24}^2\,g_{33}^2\,g_{44} +
      2\,\partial_1g_{34}\,g_{14}\,g_{23}^2\,g_{24}\,g_{34}\,g_{44} - 2\,\partial_1g_{34}\,g_{13}\,g_{23}\,g_{24}^2\,g_{34}\,g_{44} +
      \partial_1g_{33}\,g_{14}\,g_{23}\,g_{24}^2\,g_{34}\,g_{44} - \partial_1g_{33}\,g_{13}\,g_{24}^3\,g_{34}\,g_{44} +
      2\,\partial_1g_{44}\,g_{13}\,g_{22}\,g_{24}\,g_{33}\,g_{34}\,g_{44} + 2\,\partial_1g_{34}\,g_{14}\,g_{22}\,g_{24}\,g_{33}\,g_{34}\,g_{44} -
      2\,\partial_1g_{44}\,g_{12}\,g_{23}\,g_{24}\,g_{33}\,g_{34}\,g_{44} - 2\,\partial_1g_{34}\,g_{12}\,g_{24}^2\,g_{33}\,g_{34}\,g_{44} -
      \partial_1g_{44}\,g_{13}\,g_{22}\,g_{23}\,g_{34}^2\,g_{44} - 2\,\partial_1g_{34}\,g_{14}\,g_{22}\,g_{23}\,g_{34}^2\,g_{44} +
      \partial_1g_{44}\,g_{12}\,g_{23}^2\,g_{34}^2\,g_{44} + 2\,\partial_1g_{24}\,g_{14}\,g_{23}^2\,g_{34}^2\,g_{44} -
      \partial_1g_{33}\,g_{14}\,g_{22}\,g_{24}\,g_{34}^2\,g_{44} + 2\,\partial_1g_{34}\,g_{12}\,g_{23}\,g_{24}\,g_{34}^2\,g_{44} -
      2\,\partial_1g_{24}\,g_{13}\,g_{23}\,g_{24}\,g_{34}^2\,g_{44} - 2\,\partial_1g_{23}\,g_{14}\,g_{23}\,g_{24}\,g_{34}^2\,g_{44} +
      \partial_1g_{33}\,g_{12}\,g_{24}^2\,g_{34}^2\,g_{44} + 2\,\partial_1g_{23}\,g_{13}\,g_{24}^2\,g_{34}^2\,g_{44} -
      2\,\partial_1g_{24}\,g_{14}\,g_{22}\,g_{33}\,g_{34}^2\,g_{44} + 2\,\partial_1g_{24}\,g_{12}\,g_{24}\,g_{33}\,g_{34}^2\,g_{44} +
      \partial_1g_{22}\,g_{14}\,g_{24}\,g_{33}\,g_{34}^2\,g_{44} - \partial_2g_{11}\,g_{24}^2\,g_{33}\,g_{34}^2\,g_{44} +
      2\,\partial_1g_{24}\,g_{13}\,g_{22}\,g_{34}^3\,g_{44} + 2\,\partial_1g_{23}\,g_{14}\,g_{22}\,g_{34}^3\,g_{44} -
      2\,\partial_1g_{24}\,g_{12}\,g_{23}\,g_{34}^3\,g_{44} - \partial_1g_{22}\,g_{14}\,g_{23}\,g_{34}^3\,g_{44} -
      2\,\partial_1g_{23}\,g_{12}\,g_{24}\,g_{34}^3\,g_{44} - \partial_1g_{22}\,g_{13}\,g_{24}\,g_{34}^3\,g_{44} +
      2\,\partial_2g_{11}\,g_{23}\,g_{24}\,g_{34}^3\,g_{44} + \partial_1g_{22}\,g_{12}\,g_{34}^4\,g_{44} - \partial_2g_{11}\,g_{22}\,g_{34}^4\,g_{44} -
      \partial_1g_{33}\,g_{14}\,g_{23}^2\,g_{24}\,g_{44}^2 + \partial_1g_{33}\,g_{13}\,g_{23}\,g_{24}^2\,g_{44}^2 -
      2\,\partial_1g_{24}\,g_{14}\,g_{23}^2\,g_{33}\,g_{44}^2 - 2\,\partial_1g_{34}\,g_{13}\,g_{22}\,g_{24}\,g_{33}\,g_{44}^2 +
      2\,\partial_1g_{34}\,g_{12}\,g_{23}\,g_{24}\,g_{33}\,g_{44}^2 + 2\,\partial_1g_{24}\,g_{13}\,g_{23}\,g_{24}\,g_{33}\,g_{44}^2 +
      2\,\partial_1g_{23}\,g_{14}\,g_{23}\,g_{24}\,g_{33}\,g_{44}^2 - 2\,\partial_1g_{23}\,g_{13}\,g_{24}^2\,g_{33}\,g_{44}^2 +
      2\,\partial_1g_{24}\,g_{14}\,g_{22}\,g_{33}^2\,g_{44}^2 - 2\,\partial_1g_{24}\,g_{12}\,g_{24}\,g_{33}^2\,g_{44}^2 -
      \partial_1g_{22}\,g_{14}\,g_{24}\,g_{33}^2\,g_{44}^2 + \partial_2g_{11}\,g_{24}^2\,g_{33}^2\,g_{44}^2 +
      2\,\partial_1g_{34}\,g_{13}\,g_{22}\,g_{23}\,g_{34}\,g_{44}^2 + \partial_1g_{33}\,g_{14}\,g_{22}\,g_{23}\,g_{34}\,g_{44}^2 -
      2\,\partial_1g_{34}\,g_{12}\,g_{23}^2\,g_{34}\,g_{44}^2 + \partial_1g_{33}\,g_{13}\,g_{22}\,g_{24}\,g_{34}\,g_{44}^2 -
      2\,\partial_1g_{33}\,g_{12}\,g_{23}\,g_{24}\,g_{34}\,g_{44}^2 - 2\,\partial_1g_{24}\,g_{13}\,g_{22}\,g_{33}\,g_{34}\,g_{44}^2 -
      2\,\partial_1g_{23}\,g_{14}\,g_{22}\,g_{33}\,g_{34}\,g_{44}^2 + 2\,\partial_1g_{24}\,g_{12}\,g_{23}\,g_{33}\,g_{34}\,g_{44}^2 +
      \partial_1g_{22}\,g_{14}\,g_{23}\,g_{33}\,g_{34}\,g_{44}^2 + 2\,\partial_1g_{23}\,g_{12}\,g_{24}\,g_{33}\,g_{34}\,g_{44}^2 +
      \partial_1g_{22}\,g_{13}\,g_{24}\,g_{33}\,g_{34}\,g_{44}^2 - 2\,\partial_2g_{11}\,g_{23}\,g_{24}\,g_{33}\,g_{34}\,g_{44}^2 -
      2\,\partial_1g_{23}\,g_{13}\,g_{22}\,g_{34}^2\,g_{44}^2 + 2\,\partial_1g_{23}\,g_{12}\,g_{23}\,g_{34}^2\,g_{44}^2 +
      \partial_1g_{22}\,g_{13}\,g_{23}\,g_{34}^2\,g_{44}^2 - \partial_2g_{11}\,g_{23}^2\,g_{34}^2\,g_{44}^2 -
      2\,\partial_1g_{22}\,g_{12}\,g_{33}\,g_{34}^2\,g_{44}^2 + 2\,\partial_2g_{11}\,g_{22}\,g_{33}\,g_{34}^2\,g_{44}^2 -
      \partial_1g_{33}\,g_{13}\,g_{22}\,g_{23}\,g_{44}^3 + \partial_1g_{33}\,g_{12}\,g_{23}^2\,g_{44}^3 +
      2\,\partial_1g_{23}\,g_{13}\,g_{22}\,g_{33}\,g_{44}^3 - 2\,\partial_1g_{23}\,g_{12}\,g_{23}\,g_{33}\,g_{44}^3 -
      \partial_1g_{22}\,g_{13}\,g_{23}\,g_{33}\,g_{44}^3 + \partial_2g_{11}\,g_{23}^2\,g_{33}\,g_{44}^3 +
      \partial_1g_{22}\,g_{12}\,g_{33}^2\,g_{44}^3 - \partial_2g_{11}\,g_{22}\,g_{33}^2\,g_{44}^3)/
     (4\,g_{44}\,(-g_{34}^2 + g_{33}\,g_{44})\,(g_{24}^2\,g_{33} - 2\,g_{23}\,g_{24}\,g_{34} + g_{22}\,g_{34}^2 +
       g_{23}^2\,g_{44} - g_{22}\,g_{33}\,g_{44}))$

$b_{12 2} = (\partial_2g_{44}\,g_{14}\,g_{23}\,g_{24}^2\,g_{33}\,g_{34} - \partial_2g_{44}\,g_{13}\,g_{24}^3\,g_{33}\,g_{34} -
      2\,\partial_2g_{44}\,g_{14}\,g_{23}^2\,g_{24}\,g_{34}^2 + 2\,\partial_2g_{44}\,g_{13}\,g_{23}\,g_{24}^2\,g_{34}^2 +
      \partial_2g_{44}\,g_{14}\,g_{22}\,g_{23}\,g_{34}^3 - \partial_2g_{44}\,g_{13}\,g_{22}\,g_{24}\,g_{34}^3 +
      \partial_2g_{44}\,g_{14}\,g_{23}^2\,g_{24}\,g_{33}\,g_{44} - \partial_2g_{44}\,g_{13}\,g_{23}\,g_{24}^2\,g_{33}\,g_{44} -
      2\,\partial_2g_{34}\,g_{14}\,g_{23}\,g_{24}^2\,g_{33}\,g_{44} + 2\,\partial_2g_{34}\,g_{13}\,g_{24}^3\,g_{33}\,g_{44} -
      \partial_2g_{44}\,g_{14}\,g_{22}\,g_{24}\,g_{33}^2\,g_{44} + \partial_2g_{44}\,g_{12}\,g_{24}^2\,g_{33}^2\,g_{44} +
      2\,\partial_2g_{34}\,g_{14}\,g_{23}^2\,g_{24}\,g_{34}\,g_{44} - 2\,\partial_2g_{34}\,g_{13}\,g_{23}\,g_{24}^2\,g_{34}\,g_{44} +
      \partial_2g_{33}\,g_{14}\,g_{23}\,g_{24}^2\,g_{34}\,g_{44} - \partial_2g_{33}\,g_{13}\,g_{24}^3\,g_{34}\,g_{44} +
      2\,\partial_2g_{44}\,g_{13}\,g_{22}\,g_{24}\,g_{33}\,g_{34}\,g_{44} + 2\,\partial_2g_{34}\,g_{14}\,g_{22}\,g_{24}\,g_{33}\,g_{34}\,g_{44} -
      2\,\partial_2g_{44}\,g_{12}\,g_{23}\,g_{24}\,g_{33}\,g_{34}\,g_{44} - 2\,\partial_2g_{34}\,g_{12}\,g_{24}^2\,g_{33}\,g_{34}\,g_{44} -
      \partial_2g_{44}\,g_{13}\,g_{22}\,g_{23}\,g_{34}^2\,g_{44} - 2\,\partial_2g_{34}\,g_{14}\,g_{22}\,g_{23}\,g_{34}^2\,g_{44} +
      \partial_2g_{44}\,g_{12}\,g_{23}^2\,g_{34}^2\,g_{44} + 2\,\partial_2g_{24}\,g_{14}\,g_{23}^2\,g_{34}^2\,g_{44} -
      \partial_2g_{33}\,g_{14}\,g_{22}\,g_{24}\,g_{34}^2\,g_{44} + 2\,\partial_2g_{34}\,g_{12}\,g_{23}\,g_{24}\,g_{34}^2\,g_{44} -
      2\,\partial_2g_{24}\,g_{13}\,g_{23}\,g_{24}\,g_{34}^2\,g_{44} - 2\,\partial_2g_{23}\,g_{14}\,g_{23}\,g_{24}\,g_{34}^2\,g_{44} +
      \partial_2g_{33}\,g_{12}\,g_{24}^2\,g_{34}^2\,g_{44} + 2\,\partial_2g_{23}\,g_{13}\,g_{24}^2\,g_{34}^2\,g_{44} -
      2\,\partial_2g_{24}\,g_{14}\,g_{22}\,g_{33}\,g_{34}^2\,g_{44} + 2\,\partial_2g_{24}\,g_{12}\,g_{24}\,g_{33}\,g_{34}^2\,g_{44} +
      \partial_2g_{22}\,g_{14}\,g_{24}\,g_{33}\,g_{34}^2\,g_{44} + \partial_1g_{22}\,g_{24}^2\,g_{33}\,g_{34}^2\,g_{44} -
      2\,\partial_2g_{12}\,g_{24}^2\,g_{33}\,g_{34}^2\,g_{44} + 2\,\partial_2g_{24}\,g_{13}\,g_{22}\,g_{34}^3\,g_{44} +
      2\,\partial_2g_{23}\,g_{14}\,g_{22}\,g_{34}^3\,g_{44} - 2\,\partial_2g_{24}\,g_{12}\,g_{23}\,g_{34}^3\,g_{44} -
      \partial_2g_{22}\,g_{14}\,g_{23}\,g_{34}^3\,g_{44} - 2\,\partial_2g_{23}\,g_{12}\,g_{24}\,g_{34}^3\,g_{44} -
      \partial_2g_{22}\,g_{13}\,g_{24}\,g_{34}^3\,g_{44} - 2\,\partial_1g_{22}\,g_{23}\,g_{24}\,g_{34}^3\,g_{44} +
      4\,\partial_2g_{12}\,g_{23}\,g_{24}\,g_{34}^3\,g_{44} + \partial_2g_{22}\,g_{12}\,g_{34}^4\,g_{44} + \partial_1g_{22}\,g_{22}\,g_{34}^4\,g_{44} -
      2\,\partial_2g_{12}\,g_{22}\,g_{34}^4\,g_{44} - \partial_2g_{33}\,g_{14}\,g_{23}^2\,g_{24}\,g_{44}^2 +
      \partial_2g_{33}\,g_{13}\,g_{23}\,g_{24}^2\,g_{44}^2 - 2\,\partial_2g_{24}\,g_{14}\,g_{23}^2\,g_{33}\,g_{44}^2 -
      2\,\partial_2g_{34}\,g_{13}\,g_{22}\,g_{24}\,g_{33}\,g_{44}^2 + 2\,\partial_2g_{34}\,g_{12}\,g_{23}\,g_{24}\,g_{33}\,g_{44}^2 +
      2\,\partial_2g_{24}\,g_{13}\,g_{23}\,g_{24}\,g_{33}\,g_{44}^2 + 2\,\partial_2g_{23}\,g_{14}\,g_{23}\,g_{24}\,g_{33}\,g_{44}^2 -
      2\,\partial_2g_{23}\,g_{13}\,g_{24}^2\,g_{33}\,g_{44}^2 + 2\,\partial_2g_{24}\,g_{14}\,g_{22}\,g_{33}^2\,g_{44}^2 -
      2\,\partial_2g_{24}\,g_{12}\,g_{24}\,g_{33}^2\,g_{44}^2 - \partial_2g_{22}\,g_{14}\,g_{24}\,g_{33}^2\,g_{44}^2 -
      \partial_1g_{22}\,g_{24}^2\,g_{33}^2\,g_{44}^2 + 2\,\partial_2g_{12}\,g_{24}^2\,g_{33}^2\,g_{44}^2 +
      2\,\partial_2g_{34}\,g_{13}\,g_{22}\,g_{23}\,g_{34}\,g_{44}^2 + \partial_2g_{33}\,g_{14}\,g_{22}\,g_{23}\,g_{34}\,g_{44}^2 -
      2\,\partial_2g_{34}\,g_{12}\,g_{23}^2\,g_{34}\,g_{44}^2 + \partial_2g_{33}\,g_{13}\,g_{22}\,g_{24}\,g_{34}\,g_{44}^2 -
      2\,\partial_2g_{33}\,g_{12}\,g_{23}\,g_{24}\,g_{34}\,g_{44}^2 - 2\,\partial_2g_{24}\,g_{13}\,g_{22}\,g_{33}\,g_{34}\,g_{44}^2 -
      2\,\partial_2g_{23}\,g_{14}\,g_{22}\,g_{33}\,g_{34}\,g_{44}^2 + 2\,\partial_2g_{24}\,g_{12}\,g_{23}\,g_{33}\,g_{34}\,g_{44}^2 +
      \partial_2g_{22}\,g_{14}\,g_{23}\,g_{33}\,g_{34}\,g_{44}^2 + 2\,\partial_2g_{23}\,g_{12}\,g_{24}\,g_{33}\,g_{34}\,g_{44}^2 +
      \partial_2g_{22}\,g_{13}\,g_{24}\,g_{33}\,g_{34}\,g_{44}^2 + 2\,\partial_1g_{22}\,g_{23}\,g_{24}\,g_{33}\,g_{34}\,g_{44}^2 -
      4\,\partial_2g_{12}\,g_{23}\,g_{24}\,g_{33}\,g_{34}\,g_{44}^2 - 2\,\partial_2g_{23}\,g_{13}\,g_{22}\,g_{34}^2\,g_{44}^2 +
      2\,\partial_2g_{23}\,g_{12}\,g_{23}\,g_{34}^2\,g_{44}^2 + \partial_2g_{22}\,g_{13}\,g_{23}\,g_{34}^2\,g_{44}^2 +
      \partial_1g_{22}\,g_{23}^2\,g_{34}^2\,g_{44}^2 - 2\,\partial_2g_{12}\,g_{23}^2\,g_{34}^2\,g_{44}^2 -
      2\,\partial_2g_{22}\,g_{12}\,g_{33}\,g_{34}^2\,g_{44}^2 - 2\,\partial_1g_{22}\,g_{22}\,g_{33}\,g_{34}^2\,g_{44}^2 +
      4\,\partial_2g_{12}\,g_{22}\,g_{33}\,g_{34}^2\,g_{44}^2 - \partial_2g_{33}\,g_{13}\,g_{22}\,g_{23}\,g_{44}^3 +
      \partial_2g_{33}\,g_{12}\,g_{23}^2\,g_{44}^3 + 2\,\partial_2g_{23}\,g_{13}\,g_{22}\,g_{33}\,g_{44}^3 -
      2\,\partial_2g_{23}\,g_{12}\,g_{23}\,g_{33}\,g_{44}^3 - \partial_2g_{22}\,g_{13}\,g_{23}\,g_{33}\,g_{44}^3 -
      \partial_1g_{22}\,g_{23}^2\,g_{33}\,g_{44}^3 + 2\,\partial_2g_{12}\,g_{23}^2\,g_{33}\,g_{44}^3 +
      \partial_2g_{22}\,g_{12}\,g_{33}^2\,g_{44}^3 + \partial_1g_{22}\,g_{22}\,g_{33}^2\,g_{44}^3 -
      2\,\partial_2g_{12}\,g_{22}\,g_{33}^2\,g_{44}^3)/(4\,g_{44}\,(-g_{34}^2 + g_{33}\,g_{44})\,
      (g_{24}^2\,g_{33} - 2\,g_{23}\,g_{24}\,g_{34} + g_{22}\,g_{34}^2 + g_{23}^2\,g_{44} - g_{22}\,g_{33}\,g_{44}))$

$b_{12 3} = (\partial_3g_{44}\,g_{14}\,g_{23}\,g_{24}^2\,g_{33}\,g_{34} - \partial_3g_{44}\,g_{13}\,g_{24}^3\,g_{33}\,g_{34} -
      2\,\partial_3g_{44}\,g_{14}\,g_{23}^2\,g_{24}\,g_{34}^2 + 2\,\partial_3g_{44}\,g_{13}\,g_{23}\,g_{24}^2\,g_{34}^2 +
      \partial_3g_{44}\,g_{14}\,g_{22}\,g_{23}\,g_{34}^3 - \partial_3g_{44}\,g_{13}\,g_{22}\,g_{24}\,g_{34}^3 +
      \partial_3g_{44}\,g_{14}\,g_{23}^2\,g_{24}\,g_{33}\,g_{44} - \partial_3g_{44}\,g_{13}\,g_{23}\,g_{24}^2\,g_{33}\,g_{44} -
      2\,\partial_3g_{34}\,g_{14}\,g_{23}\,g_{24}^2\,g_{33}\,g_{44} + 2\,\partial_3g_{34}\,g_{13}\,g_{24}^3\,g_{33}\,g_{44} -
      \partial_3g_{44}\,g_{14}\,g_{22}\,g_{24}\,g_{33}^2\,g_{44} + \partial_3g_{44}\,g_{12}\,g_{24}^2\,g_{33}^2\,g_{44} +
      2\,\partial_3g_{34}\,g_{14}\,g_{23}^2\,g_{24}\,g_{34}\,g_{44} - 2\,\partial_3g_{34}\,g_{13}\,g_{23}\,g_{24}^2\,g_{34}\,g_{44} +
      \partial_3g_{33}\,g_{14}\,g_{23}\,g_{24}^2\,g_{34}\,g_{44} - \partial_3g_{33}\,g_{13}\,g_{24}^3\,g_{34}\,g_{44} +
      2\,\partial_3g_{44}\,g_{13}\,g_{22}\,g_{24}\,g_{33}\,g_{34}\,g_{44} + 2\,\partial_3g_{34}\,g_{14}\,g_{22}\,g_{24}\,g_{33}\,g_{34}\,g_{44} -
      2\,\partial_3g_{44}\,g_{12}\,g_{23}\,g_{24}\,g_{33}\,g_{34}\,g_{44} - 2\,\partial_3g_{34}\,g_{12}\,g_{24}^2\,g_{33}\,g_{34}\,g_{44} -
      \partial_3g_{44}\,g_{13}\,g_{22}\,g_{23}\,g_{34}^2\,g_{44} - 2\,\partial_3g_{34}\,g_{14}\,g_{22}\,g_{23}\,g_{34}^2\,g_{44} +
      \partial_3g_{44}\,g_{12}\,g_{23}^2\,g_{34}^2\,g_{44} + 2\,\partial_3g_{24}\,g_{14}\,g_{23}^2\,g_{34}^2\,g_{44} -
      \partial_3g_{33}\,g_{14}\,g_{22}\,g_{24}\,g_{34}^2\,g_{44} + 2\,\partial_3g_{34}\,g_{12}\,g_{23}\,g_{24}\,g_{34}^2\,g_{44} -
      2\,\partial_3g_{24}\,g_{13}\,g_{23}\,g_{24}\,g_{34}^2\,g_{44} - 2\,\partial_3g_{23}\,g_{14}\,g_{23}\,g_{24}\,g_{34}^2\,g_{44} +
      \partial_3g_{33}\,g_{12}\,g_{24}^2\,g_{34}^2\,g_{44} + 2\,\partial_3g_{23}\,g_{13}\,g_{24}^2\,g_{34}^2\,g_{44} -
      2\,\partial_3g_{24}\,g_{14}\,g_{22}\,g_{33}\,g_{34}^2\,g_{44} + 2\,\partial_3g_{24}\,g_{12}\,g_{24}\,g_{33}\,g_{34}^2\,g_{44} +
      \partial_3g_{22}\,g_{14}\,g_{24}\,g_{33}\,g_{34}^2\,g_{44} + \partial_1g_{23}\,g_{24}^2\,g_{33}\,g_{34}^2\,g_{44} -
      \partial_2g_{13}\,g_{24}^2\,g_{33}\,g_{34}^2\,g_{44} - \partial_3g_{12}\,g_{24}^2\,g_{33}\,g_{34}^2\,g_{44} +
      2\,\partial_3g_{24}\,g_{13}\,g_{22}\,g_{34}^3\,g_{44} + 2\,\partial_3g_{23}\,g_{14}\,g_{22}\,g_{34}^3\,g_{44} -
      2\,\partial_3g_{24}\,g_{12}\,g_{23}\,g_{34}^3\,g_{44} - \partial_3g_{22}\,g_{14}\,g_{23}\,g_{34}^3\,g_{44} -
      2\,\partial_3g_{23}\,g_{12}\,g_{24}\,g_{34}^3\,g_{44} - \partial_3g_{22}\,g_{13}\,g_{24}\,g_{34}^3\,g_{44} -
      2\,\partial_1g_{23}\,g_{23}\,g_{24}\,g_{34}^3\,g_{44} + 2\,\partial_2g_{13}\,g_{23}\,g_{24}\,g_{34}^3\,g_{44} +
      2\,\partial_3g_{12}\,g_{23}\,g_{24}\,g_{34}^3\,g_{44} + \partial_3g_{22}\,g_{12}\,g_{34}^4\,g_{44} + \partial_1g_{23}\,g_{22}\,g_{34}^4\,g_{44} -
      \partial_2g_{13}\,g_{22}\,g_{34}^4\,g_{44} - \partial_3g_{12}\,g_{22}\,g_{34}^4\,g_{44} - \partial_3g_{33}\,g_{14}\,g_{23}^2\,g_{24}\,g_{44}^2 +
      \partial_3g_{33}\,g_{13}\,g_{23}\,g_{24}^2\,g_{44}^2 - 2\,\partial_3g_{24}\,g_{14}\,g_{23}^2\,g_{33}\,g_{44}^2 -
      2\,\partial_3g_{34}\,g_{13}\,g_{22}\,g_{24}\,g_{33}\,g_{44}^2 + 2\,\partial_3g_{34}\,g_{12}\,g_{23}\,g_{24}\,g_{33}\,g_{44}^2 +
      2\,\partial_3g_{24}\,g_{13}\,g_{23}\,g_{24}\,g_{33}\,g_{44}^2 + 2\,\partial_3g_{23}\,g_{14}\,g_{23}\,g_{24}\,g_{33}\,g_{44}^2 -
      2\,\partial_3g_{23}\,g_{13}\,g_{24}^2\,g_{33}\,g_{44}^2 + 2\,\partial_3g_{24}\,g_{14}\,g_{22}\,g_{33}^2\,g_{44}^2 -
      2\,\partial_3g_{24}\,g_{12}\,g_{24}\,g_{33}^2\,g_{44}^2 - \partial_3g_{22}\,g_{14}\,g_{24}\,g_{33}^2\,g_{44}^2 -
      \partial_1g_{23}\,g_{24}^2\,g_{33}^2\,g_{44}^2 + \partial_2g_{13}\,g_{24}^2\,g_{33}^2\,g_{44}^2 +
      \partial_3g_{12}\,g_{24}^2\,g_{33}^2\,g_{44}^2 + 2\,\partial_3g_{34}\,g_{13}\,g_{22}\,g_{23}\,g_{34}\,g_{44}^2 +
      \partial_3g_{33}\,g_{14}\,g_{22}\,g_{23}\,g_{34}\,g_{44}^2 - 2\,\partial_3g_{34}\,g_{12}\,g_{23}^2\,g_{34}\,g_{44}^2 +
      \partial_3g_{33}\,g_{13}\,g_{22}\,g_{24}\,g_{34}\,g_{44}^2 - 2\,\partial_3g_{33}\,g_{12}\,g_{23}\,g_{24}\,g_{34}\,g_{44}^2 -
      2\,\partial_3g_{24}\,g_{13}\,g_{22}\,g_{33}\,g_{34}\,g_{44}^2 - 2\,\partial_3g_{23}\,g_{14}\,g_{22}\,g_{33}\,g_{34}\,g_{44}^2 +
      2\,\partial_3g_{24}\,g_{12}\,g_{23}\,g_{33}\,g_{34}\,g_{44}^2 + \partial_3g_{22}\,g_{14}\,g_{23}\,g_{33}\,g_{34}\,g_{44}^2 +
      2\,\partial_3g_{23}\,g_{12}\,g_{24}\,g_{33}\,g_{34}\,g_{44}^2 + \partial_3g_{22}\,g_{13}\,g_{24}\,g_{33}\,g_{34}\,g_{44}^2 +
      2\,\partial_1g_{23}\,g_{23}\,g_{24}\,g_{33}\,g_{34}\,g_{44}^2 - 2\,\partial_2g_{13}\,g_{23}\,g_{24}\,g_{33}\,g_{34}\,g_{44}^2 -
      2\,\partial_3g_{12}\,g_{23}\,g_{24}\,g_{33}\,g_{34}\,g_{44}^2 - 2\,\partial_3g_{23}\,g_{13}\,g_{22}\,g_{34}^2\,g_{44}^2 +
      2\,\partial_3g_{23}\,g_{12}\,g_{23}\,g_{34}^2\,g_{44}^2 + \partial_3g_{22}\,g_{13}\,g_{23}\,g_{34}^2\,g_{44}^2 +
      \partial_1g_{23}\,g_{23}^2\,g_{34}^2\,g_{44}^2 - \partial_2g_{13}\,g_{23}^2\,g_{34}^2\,g_{44}^2 -
      \partial_3g_{12}\,g_{23}^2\,g_{34}^2\,g_{44}^2 - 2\,\partial_3g_{22}\,g_{12}\,g_{33}\,g_{34}^2\,g_{44}^2 -
      2\,\partial_1g_{23}\,g_{22}\,g_{33}\,g_{34}^2\,g_{44}^2 + 2\,\partial_2g_{13}\,g_{22}\,g_{33}\,g_{34}^2\,g_{44}^2 +
      2\,\partial_3g_{12}\,g_{22}\,g_{33}\,g_{34}^2\,g_{44}^2 - \partial_3g_{33}\,g_{13}\,g_{22}\,g_{23}\,g_{44}^3 +
      \partial_3g_{33}\,g_{12}\,g_{23}^2\,g_{44}^3 + 2\,\partial_3g_{23}\,g_{13}\,g_{22}\,g_{33}\,g_{44}^3 -
      2\,\partial_3g_{23}\,g_{12}\,g_{23}\,g_{33}\,g_{44}^3 - \partial_3g_{22}\,g_{13}\,g_{23}\,g_{33}\,g_{44}^3 -
      \partial_1g_{23}\,g_{23}^2\,g_{33}\,g_{44}^3 + \partial_2g_{13}\,g_{23}^2\,g_{33}\,g_{44}^3 + \partial_3g_{12}\,g_{23}^2\,g_{33}\,g_{44}^3 +
      \partial_3g_{22}\,g_{12}\,g_{33}^2\,g_{44}^3 + \partial_1g_{23}\,g_{22}\,g_{33}^2\,g_{44}^3 - \partial_2g_{13}\,g_{22}\,g_{33}^2\,g_{44}^3 -
      \partial_3g_{12}\,g_{22}\,g_{33}^2\,g_{44}^3)/(4\,g_{44}\,(-g_{34}^2 + g_{33}\,g_{44})\,
      (g_{24}^2\,g_{33} - 2\,g_{23}\,g_{24}\,g_{34} + g_{22}\,g_{34}^2 + g_{23}^2\,g_{44} - g_{22}\,g_{33}\,g_{44}))$

$b_{12 4} = (\partial_4g_{44}\,g_{14}\,g_{23}\,g_{24}^2\,g_{33}\,g_{34} - \partial_4g_{44}\,g_{13}\,g_{24}^3\,g_{33}\,g_{34} -
      2\,\partial_4g_{44}\,g_{14}\,g_{23}^2\,g_{24}\,g_{34}^2 + 2\,\partial_4g_{44}\,g_{13}\,g_{23}\,g_{24}^2\,g_{34}^2 +
      \partial_4g_{44}\,g_{14}\,g_{22}\,g_{23}\,g_{34}^3 - \partial_4g_{44}\,g_{13}\,g_{22}\,g_{24}\,g_{34}^3 +
      \partial_4g_{44}\,g_{14}\,g_{23}^2\,g_{24}\,g_{33}\,g_{44} - \partial_4g_{44}\,g_{13}\,g_{23}\,g_{24}^2\,g_{33}\,g_{44} -
      2\,\partial_4g_{34}\,g_{14}\,g_{23}\,g_{24}^2\,g_{33}\,g_{44} + 2\,\partial_4g_{34}\,g_{13}\,g_{24}^3\,g_{33}\,g_{44} -
      \partial_4g_{44}\,g_{14}\,g_{22}\,g_{24}\,g_{33}^2\,g_{44} + \partial_4g_{44}\,g_{12}\,g_{24}^2\,g_{33}^2\,g_{44} +
      2\,\partial_4g_{34}\,g_{14}\,g_{23}^2\,g_{24}\,g_{34}\,g_{44} - 2\,\partial_4g_{34}\,g_{13}\,g_{23}\,g_{24}^2\,g_{34}\,g_{44} +
      \partial_4g_{33}\,g_{14}\,g_{23}\,g_{24}^2\,g_{34}\,g_{44} - \partial_4g_{33}\,g_{13}\,g_{24}^3\,g_{34}\,g_{44} +
      2\,\partial_4g_{44}\,g_{13}\,g_{22}\,g_{24}\,g_{33}\,g_{34}\,g_{44} + 2\,\partial_4g_{34}\,g_{14}\,g_{22}\,g_{24}\,g_{33}\,g_{34}\,g_{44} -
      2\,\partial_4g_{44}\,g_{12}\,g_{23}\,g_{24}\,g_{33}\,g_{34}\,g_{44} - 2\,\partial_4g_{34}\,g_{12}\,g_{24}^2\,g_{33}\,g_{34}\,g_{44} -
      \partial_4g_{44}\,g_{13}\,g_{22}\,g_{23}\,g_{34}^2\,g_{44} - 2\,\partial_4g_{34}\,g_{14}\,g_{22}\,g_{23}\,g_{34}^2\,g_{44} +
      \partial_4g_{44}\,g_{12}\,g_{23}^2\,g_{34}^2\,g_{44} + 2\,\partial_4g_{24}\,g_{14}\,g_{23}^2\,g_{34}^2\,g_{44} -
      \partial_4g_{33}\,g_{14}\,g_{22}\,g_{24}\,g_{34}^2\,g_{44} + 2\,\partial_4g_{34}\,g_{12}\,g_{23}\,g_{24}\,g_{34}^2\,g_{44} -
      2\,\partial_4g_{24}\,g_{13}\,g_{23}\,g_{24}\,g_{34}^2\,g_{44} - 2\,\partial_4g_{23}\,g_{14}\,g_{23}\,g_{24}\,g_{34}^2\,g_{44} +
      \partial_4g_{33}\,g_{12}\,g_{24}^2\,g_{34}^2\,g_{44} + 2\,\partial_4g_{23}\,g_{13}\,g_{24}^2\,g_{34}^2\,g_{44} -
      2\,\partial_4g_{24}\,g_{14}\,g_{22}\,g_{33}\,g_{34}^2\,g_{44} + 2\,\partial_4g_{24}\,g_{12}\,g_{24}\,g_{33}\,g_{34}^2\,g_{44} +
      \partial_4g_{22}\,g_{14}\,g_{24}\,g_{33}\,g_{34}^2\,g_{44} + \partial_1g_{24}\,g_{24}^2\,g_{33}\,g_{34}^2\,g_{44} -
      \partial_2g_{14}\,g_{24}^2\,g_{33}\,g_{34}^2\,g_{44} - \partial_4g_{12}\,g_{24}^2\,g_{33}\,g_{34}^2\,g_{44} +
      2\,\partial_4g_{24}\,g_{13}\,g_{22}\,g_{34}^3\,g_{44} + 2\,\partial_4g_{23}\,g_{14}\,g_{22}\,g_{34}^3\,g_{44} -
      2\,\partial_4g_{24}\,g_{12}\,g_{23}\,g_{34}^3\,g_{44} - \partial_4g_{22}\,g_{14}\,g_{23}\,g_{34}^3\,g_{44} -
      2\,\partial_4g_{23}\,g_{12}\,g_{24}\,g_{34}^3\,g_{44} - \partial_4g_{22}\,g_{13}\,g_{24}\,g_{34}^3\,g_{44} -
      2\,\partial_1g_{24}\,g_{23}\,g_{24}\,g_{34}^3\,g_{44} + 2\,\partial_2g_{14}\,g_{23}\,g_{24}\,g_{34}^3\,g_{44} +
      2\,\partial_4g_{12}\,g_{23}\,g_{24}\,g_{34}^3\,g_{44} + \partial_4g_{22}\,g_{12}\,g_{34}^4\,g_{44} + \partial_1g_{24}\,g_{22}\,g_{34}^4\,g_{44} -
      \partial_2g_{14}\,g_{22}\,g_{34}^4\,g_{44} - \partial_4g_{12}\,g_{22}\,g_{34}^4\,g_{44} - \partial_4g_{33}\,g_{14}\,g_{23}^2\,g_{24}\,g_{44}^2 +
      \partial_4g_{33}\,g_{13}\,g_{23}\,g_{24}^2\,g_{44}^2 - 2\,\partial_4g_{24}\,g_{14}\,g_{23}^2\,g_{33}\,g_{44}^2 -
      2\,\partial_4g_{34}\,g_{13}\,g_{22}\,g_{24}\,g_{33}\,g_{44}^2 + 2\,\partial_4g_{34}\,g_{12}\,g_{23}\,g_{24}\,g_{33}\,g_{44}^2 +
      2\,\partial_4g_{24}\,g_{13}\,g_{23}\,g_{24}\,g_{33}\,g_{44}^2 + 2\,\partial_4g_{23}\,g_{14}\,g_{23}\,g_{24}\,g_{33}\,g_{44}^2 -
      2\,\partial_4g_{23}\,g_{13}\,g_{24}^2\,g_{33}\,g_{44}^2 + 2\,\partial_4g_{24}\,g_{14}\,g_{22}\,g_{33}^2\,g_{44}^2 -
      2\,\partial_4g_{24}\,g_{12}\,g_{24}\,g_{33}^2\,g_{44}^2 - \partial_4g_{22}\,g_{14}\,g_{24}\,g_{33}^2\,g_{44}^2 -
      \partial_1g_{24}\,g_{24}^2\,g_{33}^2\,g_{44}^2 + \partial_2g_{14}\,g_{24}^2\,g_{33}^2\,g_{44}^2 +
      \partial_4g_{12}\,g_{24}^2\,g_{33}^2\,g_{44}^2 + 2\,\partial_4g_{34}\,g_{13}\,g_{22}\,g_{23}\,g_{34}\,g_{44}^2 +
      \partial_4g_{33}\,g_{14}\,g_{22}\,g_{23}\,g_{34}\,g_{44}^2 - 2\,\partial_4g_{34}\,g_{12}\,g_{23}^2\,g_{34}\,g_{44}^2 +
      \partial_4g_{33}\,g_{13}\,g_{22}\,g_{24}\,g_{34}\,g_{44}^2 - 2\,\partial_4g_{33}\,g_{12}\,g_{23}\,g_{24}\,g_{34}\,g_{44}^2 -
      2\,\partial_4g_{24}\,g_{13}\,g_{22}\,g_{33}\,g_{34}\,g_{44}^2 - 2\,\partial_4g_{23}\,g_{14}\,g_{22}\,g_{33}\,g_{34}\,g_{44}^2 +
      2\,\partial_4g_{24}\,g_{12}\,g_{23}\,g_{33}\,g_{34}\,g_{44}^2 + \partial_4g_{22}\,g_{14}\,g_{23}\,g_{33}\,g_{34}\,g_{44}^2 +
      2\,\partial_4g_{23}\,g_{12}\,g_{24}\,g_{33}\,g_{34}\,g_{44}^2 + \partial_4g_{22}\,g_{13}\,g_{24}\,g_{33}\,g_{34}\,g_{44}^2 +
      2\,\partial_1g_{24}\,g_{23}\,g_{24}\,g_{33}\,g_{34}\,g_{44}^2 - 2\,\partial_2g_{14}\,g_{23}\,g_{24}\,g_{33}\,g_{34}\,g_{44}^2 -
      2\,\partial_4g_{12}\,g_{23}\,g_{24}\,g_{33}\,g_{34}\,g_{44}^2 - 2\,\partial_4g_{23}\,g_{13}\,g_{22}\,g_{34}^2\,g_{44}^2 +
      2\,\partial_4g_{23}\,g_{12}\,g_{23}\,g_{34}^2\,g_{44}^2 + \partial_4g_{22}\,g_{13}\,g_{23}\,g_{34}^2\,g_{44}^2 +
      \partial_1g_{24}\,g_{23}^2\,g_{34}^2\,g_{44}^2 - \partial_2g_{14}\,g_{23}^2\,g_{34}^2\,g_{44}^2 -
      \partial_4g_{12}\,g_{23}^2\,g_{34}^2\,g_{44}^2 - 2\,\partial_4g_{22}\,g_{12}\,g_{33}\,g_{34}^2\,g_{44}^2 -
      2\,\partial_1g_{24}\,g_{22}\,g_{33}\,g_{34}^2\,g_{44}^2 + 2\,\partial_2g_{14}\,g_{22}\,g_{33}\,g_{34}^2\,g_{44}^2 +
      2\,\partial_4g_{12}\,g_{22}\,g_{33}\,g_{34}^2\,g_{44}^2 - \partial_4g_{33}\,g_{13}\,g_{22}\,g_{23}\,g_{44}^3 +
      \partial_4g_{33}\,g_{12}\,g_{23}^2\,g_{44}^3 + 2\,\partial_4g_{23}\,g_{13}\,g_{22}\,g_{33}\,g_{44}^3 -
      2\,\partial_4g_{23}\,g_{12}\,g_{23}\,g_{33}\,g_{44}^3 - \partial_4g_{22}\,g_{13}\,g_{23}\,g_{33}\,g_{44}^3 -
      \partial_1g_{24}\,g_{23}^2\,g_{33}\,g_{44}^3 + \partial_2g_{14}\,g_{23}^2\,g_{33}\,g_{44}^3 + \partial_4g_{12}\,g_{23}^2\,g_{33}\,g_{44}^3 +
      \partial_4g_{22}\,g_{12}\,g_{33}^2\,g_{44}^3 + \partial_1g_{24}\,g_{22}\,g_{33}^2\,g_{44}^3 - \partial_2g_{14}\,g_{22}\,g_{33}^2\,g_{44}^3 -
      \partial_4g_{12}\,g_{22}\,g_{33}^2\,g_{44}^3)/(4\,g_{44}\,(-g_{34}^2 + g_{33}\,g_{44})\,
      (g_{24}^2\,g_{33} - 2\,g_{23}\,g_{24}\,g_{34} + g_{22}\,g_{34}^2 + g_{23}^2\,g_{44} - g_{22}\,g_{33}\,g_{44}))$

$b_{13 1} = (\partial_1g_{44}\,g_{14}\,g_{33}\,g_{34} - \partial_1g_{44}\,g_{13}\,g_{34}^2 - 2\,\partial_1g_{34}\,g_{14}\,g_{33}\,g_{44} +
      2\,\partial_1g_{34}\,g_{13}\,g_{34}\,g_{44} + \partial_1g_{33}\,g_{14}\,g_{34}\,g_{44} - \partial_3g_{11}\,g_{34}^2\,g_{44} -
      \partial_1g_{33}\,g_{13}\,g_{44}^2 + \partial_3g_{11}\,g_{33}\,g_{44}^2)/(4\,g_{44}\,(-g_{34}^2 + g_{33}\,g_{44}))$

$b_{13 2} = (\partial_2g_{44}\,g_{14}\,g_{33}\,g_{34} - \partial_2g_{44}\,g_{13}\,g_{34}^2 - 2\,\partial_2g_{34}\,g_{14}\,g_{33}\,g_{44} +
      2\,\partial_2g_{34}\,g_{13}\,g_{34}\,g_{44} + \partial_2g_{33}\,g_{14}\,g_{34}\,g_{44} + \partial_1g_{23}\,g_{34}^2\,g_{44} -
      \partial_2g_{13}\,g_{34}^2\,g_{44} - \partial_3g_{12}\,g_{34}^2\,g_{44} - \partial_2g_{33}\,g_{13}\,g_{44}^2 - \partial_1g_{23}\,g_{33}\,g_{44}^2 +
      \partial_2g_{13}\,g_{33}\,g_{44}^2 + \partial_3g_{12}\,g_{33}\,g_{44}^2)/(4\,g_{44}\,(-g_{34}^2 + g_{33}\,g_{44}))$

$b_{13 3} = (\partial_3g_{44}\,g_{14}\,g_{33}\,g_{34} - \partial_3g_{44}\,g_{13}\,g_{34}^2 - 2\,\partial_3g_{34}\,g_{14}\,g_{33}\,g_{44} +
      2\,\partial_3g_{34}\,g_{13}\,g_{34}\,g_{44} + \partial_3g_{33}\,g_{14}\,g_{34}\,g_{44} + \partial_1g_{33}\,g_{34}^2\,g_{44} -
      2\,\partial_3g_{13}\,g_{34}^2\,g_{44} - \partial_3g_{33}\,g_{13}\,g_{44}^2 - \partial_1g_{33}\,g_{33}\,g_{44}^2 +
      2\,\partial_3g_{13}\,g_{33}\,g_{44}^2)/(4\,g_{44}\,(-g_{34}^2 + g_{33}\,g_{44}))$

$b_{13 4} = (\partial_4g_{44}\,g_{14}\,g_{33}\,g_{34} - \partial_4g_{44}\,g_{13}\,g_{34}^2 - 2\,\partial_4g_{34}\,g_{14}\,g_{33}\,g_{44} +
      2\,\partial_4g_{34}\,g_{13}\,g_{34}\,g_{44} + \partial_4g_{33}\,g_{14}\,g_{34}\,g_{44} + \partial_1g_{34}\,g_{34}^2\,g_{44} -
      \partial_3g_{14}\,g_{34}^2\,g_{44} - \partial_4g_{13}\,g_{34}^2\,g_{44} - \partial_4g_{33}\,g_{13}\,g_{44}^2 - \partial_1g_{34}\,g_{33}\,g_{44}^2 +
      \partial_3g_{14}\,g_{33}\,g_{44}^2 + \partial_4g_{13}\,g_{33}\,g_{44}^2)/(4\,g_{44}\,(-g_{34}^2 + g_{33}\,g_{44}))$

$b_{14 1} = (-(\partial_1g_{44}\,g_{14}) + \partial_4g_{11}\,g_{44})/(4\,g_{44})$

$b_{14 2} = (-(\partial_2g_{44}\,g_{14}) - \partial_1g_{24}\,g_{44} + \partial_2g_{14}\,g_{44} + \partial_4g_{12}\,g_{44})/(4\,g_{44})$

$b_{14 3} = (-(\partial_3g_{44}\,g_{14}) - \partial_1g_{34}\,g_{44} + \partial_3g_{14}\,g_{44} + \partial_4g_{13}\,g_{44})/(4\,g_{44})$

$b_{14 4} = (-(\partial_4g_{44}\,g_{14}) - \partial_1g_{44}\,g_{44} + 2\,\partial_4g_{14}\,g_{44})/(4\,g_{44})$

$b_{23 1} = (-(\partial_1g_{44}\,g_{24}\,g_{33}\,g_{34}) + \partial_1g_{44}\,g_{23}\,g_{34}^2 + 2\,\partial_1g_{34}\,g_{24}\,g_{33}\,g_{44} -
      2\,\partial_1g_{34}\,g_{23}\,g_{34}\,g_{44} - \partial_1g_{33}\,g_{24}\,g_{34}\,g_{44} + \partial_1g_{23}\,g_{34}^2\,g_{44} -
      \partial_2g_{13}\,g_{34}^2\,g_{44} + \partial_3g_{12}\,g_{34}^2\,g_{44} + \partial_1g_{33}\,g_{23}\,g_{44}^2 - \partial_1g_{23}\,g_{33}\,g_{44}^2 +
      \partial_2g_{13}\,g_{33}\,g_{44}^2 - \partial_3g_{12}\,g_{33}\,g_{44}^2)/(4\,g_{44}\,(g_{34}^2 - g_{33}\,g_{44}))$

$b_{23 2} = (-(\partial_2g_{44}\,g_{24}\,g_{33}\,g_{34}) + \partial_2g_{44}\,g_{23}\,g_{34}^2 + 2\,\partial_2g_{34}\,g_{24}\,g_{33}\,g_{44} -
      2\,\partial_2g_{34}\,g_{23}\,g_{34}\,g_{44} - \partial_2g_{33}\,g_{24}\,g_{34}\,g_{44} + \partial_3g_{22}\,g_{34}^2\,g_{44} +
      \partial_2g_{33}\,g_{23}\,g_{44}^2 - \partial_3g_{22}\,g_{33}\,g_{44}^2)/(4\,g_{44}\,(g_{34}^2 - g_{33}\,g_{44}))$

$b_{23 3} = (-(\partial_3g_{44}\,g_{24}\,g_{33}\,g_{34}) + \partial_3g_{44}\,g_{23}\,g_{34}^2 + 2\,\partial_3g_{34}\,g_{24}\,g_{33}\,g_{44} -
      2\,\partial_3g_{34}\,g_{23}\,g_{34}\,g_{44} - \partial_3g_{33}\,g_{24}\,g_{34}\,g_{44} - \partial_2g_{33}\,g_{34}^2\,g_{44} +
      2\,\partial_3g_{23}\,g_{34}^2\,g_{44} + \partial_3g_{33}\,g_{23}\,g_{44}^2 + \partial_2g_{33}\,g_{33}\,g_{44}^2 -
      2\,\partial_3g_{23}\,g_{33}\,g_{44}^2)/(4\,g_{44}\,(g_{34}^2 - g_{33}\,g_{44}))$

$b_{23 4} = (-(\partial_4g_{44}\,g_{24}\,g_{33}\,g_{34}) + \partial_4g_{44}\,g_{23}\,g_{34}^2 + 2\,\partial_4g_{34}\,g_{24}\,g_{33}\,g_{44} -
      2\,\partial_4g_{34}\,g_{23}\,g_{34}\,g_{44} - \partial_4g_{33}\,g_{24}\,g_{34}\,g_{44} - \partial_2g_{34}\,g_{34}^2\,g_{44} +
      \partial_3g_{24}\,g_{34}^2\,g_{44} + \partial_4g_{23}\,g_{34}^2\,g_{44} + \partial_4g_{33}\,g_{23}\,g_{44}^2 + \partial_2g_{34}\,g_{33}\,g_{44}^2 -
      \partial_3g_{24}\,g_{33}\,g_{44}^2 - \partial_4g_{23}\,g_{33}\,g_{44}^2)/(4\,g_{44}\,(g_{34}^2 - g_{33}\,g_{44}))$

$b_{24 1} = (-(\partial_1g_{44}\,g_{24}) + \partial_1g_{24}\,g_{44} - \partial_2g_{14}\,g_{44} + \partial_4g_{12}\,g_{44})/(4\,g_{44})$

$b_{24 2} = (-(\partial_2g_{44}\,g_{24}) + \partial_4g_{22}\,g_{44})/(4\,g_{44})$

$b_{24 3} = (-(\partial_3g_{44}\,g_{24}) - \partial_2g_{34}\,g_{44} + \partial_3g_{24}\,g_{44} + \partial_4g_{23}\,g_{44})/(4\,g_{44})$

$b_{24 4} = (-(\partial_4g_{44}\,g_{24}) - \partial_2g_{44}\,g_{44} + 2\,\partial_4g_{24}\,g_{44})/(4\,g_{44})$

$b_{34 1} = (-(\partial_1g_{44}\,g_{34}) + \partial_1g_{34}\,g_{44} - \partial_3g_{14}\,g_{44} + \partial_4g_{13}\,g_{44})/(4\,g_{44})$

$b_{34 2} = (-(\partial_2g_{44}\,g_{34}) + \partial_2g_{34}\,g_{44} - \partial_3g_{24}\,g_{44} + \partial_4g_{23}\,g_{44})/(4\,g_{44})$

$b_{34 3} = (-(\partial_3g_{44}\,g_{34}) + \partial_4g_{33}\,g_{44})/(4\,g_{44})$

$b_{34 4} = (-(\partial_4g_{44}\,g_{34}) - \partial_3g_{44}\,g_{44} +
2\,\partial_4g_{34}\,g_{44})/(4\,g_{44})$.

\medskip

Here are the formulas for $b_{\alpha\beta\mu}$ in the temporal gauge
$(g_{11}=1,\,g_{12}=g_{13}=g_{14}=0)$

\medskip

$b_{12 1} = 0$

$b_{12 2} = -\partial_{1}g_{22}/4$

$b_{12 3} = -\partial_{1}g_{23}/4$

$b_{12 4} = -\partial_{1}g_{24}/4$

$b_{13 1} = 0$

$b_{13 2} = -\partial_{1}g_{23}/4$

$b_{13 3} = -\partial_{1}g_{33}/4$

$b_{13 4} = -\partial_{1}g_{34}/4$

$b_{14 1} = 0$

$b_{14 2} = -\partial_{1}g_{24}/4$

$b_{14 3} = -\partial_{1}g_{34}/4$

$b_{14 4} = -\partial_{1}g_{44}/4$

$b_{23 1} = -(\partial_{1}g_{44}\,g_{24}\,g_{33}\,g_{34} - \partial_{1}g_{44}\,g_{23}\,g_{34}^2 - 2\,\partial_{1}g_{34}\,g_{24}\,g_{33}\,g_{44} +
      2\,\partial_{1}g_{34}\,g_{23}\,g_{34}\,g_{44}
      + \partial_{1}g_{33}\,g_{24}\,g_{34}\,g_{44} - \partial_{1}g_{23}\,g_{34}^2\,g_{44} -
      \partial_{1}g_{33}\,g_{23}\,g_{44}^2 + \partial_{1}g_{23}\,g_{33}\,g_{44}^2)/(4\,g_{44}\,(g_{34}^2 - g_{33}\,g_{44}))$

$b_{23 2} = -(\partial_{2}g_{44}\,g_{24}\,g_{33}\,g_{34} - \partial_{2}g_{44}\,g_{23}\,g_{34}^2 - 2\,\partial_{2}g_{34}\,g_{24}\,g_{33}\,g_{44} +
      2\,\partial_{2}g_{34}\,g_{23}\,g_{34}\,g_{44}
      + \partial_{2}g_{33}\,g_{24}\,g_{34}\,g_{44} - \partial_{3}g_{22}\,g_{34}^2\,g_{44} -
      \partial_{2}g_{33}\,g_{23}\,g_{44}^2 + \partial_{3}g_{22}\,g_{33}\,g_{44}^2)/(4\,g_{44}\,(g_{34}^2 - g_{33}\,g_{44}))$

$b_{23 3} = -((\partial_{3}g_{44}\,g_{24}\,g_{33}\,g_{34} - \partial_{3}g_{44}\,g_{23}\,g_{34}^2 - 2\,\partial_{3}g_{34}\,g_{24}\,g_{33}\,g_{44} +
       2\,\partial_{3}g_{34}\,g_{23}\,g_{34}\,g_{44}
       + \partial_{3}g_{33}\,g_{24}\,g_{34}\,g_{44} + \partial_{2}g_{33}\,g_{34}^2\,g_{44} -
       2\,\partial_{3}g_{23}\,g_{34}^2\,g_{44}-
       \partial_{3}g_{33}\,g_{23}\,g_{44}^2
        - \partial_{2}g_{33}\,g_{33}\,g_{44}^2 +
       2\,\partial_{3}g_{23}\,g_{33}\,g_{44}^2)/(4\,g_{34}^2\,g_{44} - 4\,g_{33}\,g_{44}^2))$

$b_{23 4} = -((\partial_{4}g_{44}\,g_{24}\,g_{33}\,g_{34} - \partial_{4}g_{44}\,g_{23}\,g_{34}^2 - 2\,\partial_{4}g_{34}\,g_{24}\,g_{33}\,g_{44} +
       2\,\partial_{4}g_{34}\,g_{23}\,g_{34}\,g_{44}
       + \partial_{4}g_{33}\,g_{24}\,g_{34}\,g_{44} + \partial_{2}g_{34}\,g_{34}^2\,g_{44} -
       \partial_{3}g_{24}\,g_{34}^2\,g_{44}-
       \partial_{4}g_{23}\,g_{34}^2\,g_{44} -
       \partial_{4}g_{33}\,g_{23}\,g_{44}^2
       -\partial_{2}g_{34}\,g_{33}\,g_{44}^2 + \partial_{3}g_{24}\,g_{33}\,g_{44}^2 + \partial_{4}g_{23}\,g_{33}\,g_{44}^2)/
     (4\,g_{34}^2\,g_{44} - 4\,g_{33}\,g_{44}^2))$

$b_{24 1} = (-(\partial_{1}g_{44}\,g_{24}) + \partial_{1}g_{24}\,g_{44})/(4\,g_{44})$

$b_{24 2} = (-(\partial_{2}g_{44}\,g_{24}) + \partial_{4}g_{22}\,g_{44})/(4\,g_{44})$

$b_{24 3} = (-(\partial_{3}g_{44}\,g_{24}) - \partial_{2}g_{34}\,g_{44} + \partial_{3}g_{24}\,g_{44} + \partial_{4}g_{23}\,g_{44})/(4\,g_{44})$

$b_{24 4} = (-(\partial_{4}g_{44}\,g_{24}) - \partial_{2}g_{44}\,g_{44} + 2\,\partial_{4}g_{24}\,g_{44})/(4\,g_{44})$

$b_{34 1} = -(\partial_{1}g_{44}\,g_{34} - \partial_{1}g_{34}\,g_{44})/(4\,g_{44})$

$b_{34 2} = -(\partial_{2}g_{44}\,g_{34} - \partial_{2}g_{34}\,g_{44} + \partial_{3}g_{24}\,g_{44} - \partial_{4}g_{23}\,g_{44})/(4\,g_{44})$

$b_{34 3} = -(\partial_{3}g_{44}\,g_{34} - \partial_{4}g_{33}\,g_{44})/(4\,g_{44})$

$b_{34 4} = -(\partial_{4}g_{44}\,g_{34} + \partial_{3}g_{44}\,g_{44} -
2\,\partial_{4}g_{34}\,g_{44})/(4\,g_{44})$.

\medskip

Here are the formulas for $b_{\alpha\beta\mu}$ in case of the diagonal metric
tensor $g_{\mu\nu}=\diag(g_{11},g_{22},g_{33},g_{44})$. These formulas
are also valid for the space dimensions $n=2,3$ (see section 8)
$$
b_{\alpha\beta\mu}=-\frac{1}{2}\partial_{[\alpha}g_{\beta]\mu}.
$$



\begin{thebibliography}{99}
\bibitem{Marchuk:AACA} Marchuk N.G., Advances in Applied Clifford Algebras, v.8, N.1,
(1998), p.181-225.(http://xxx.lanl.gov/abs/math-ph/9811022)\par
\bibitem{Marchuk:Cimento} Marchuk N.G., Nuovo Cimento, 115B, 11, 2000.\par
\bibitem{Marchuk:Cimento1} Marchuk N.G., Dirac type tensor equations,
to appear in Nuovo Cimento B in 2001.\par
\bibitem{Dirac} Dirac P.A.M., Proc. Roy. Soc. Lond. A117 (1928) 610.\par
\bibitem{Ivanenko} Ivanenko D., Landau L., Z. Phys., 48 (1928)340.\par
\bibitem{Kahler} K\"ahler E., Randiconti di Mat. (Roma) ser. 5, 21, (1962) 425.\par
\bibitem{Gursey} G\"ursey F., Nuovo Cimento, 3, p.988 (1956).\par
\bibitem{Hestenes} Hestenes D., {\sl Space-Time Algebra},
Gordon and Breach, New York, 1966.\par
\bibitem{Hestenes1} Hestenes D., J. Math. Phys., 8, pp.798-808
(1967).\par
\bibitem{Nakahara} Nakahara M., {\sl Geometry, Topology and Physics,}
Institute of Physics Publishing, Bristol and Philadelphia, 1998.\par
\end{thebibliography}
\end{document}